\documentclass[journal]{IEEEtran}
\usepackage{amsmath,amsfonts}
\usepackage{algorithmic}
\usepackage{algorithm}
\usepackage{array}
\usepackage[caption=false,font=normalsize,labelfont=sf,textfont=sf]{subfig}
\usepackage{textcomp}
\usepackage{stfloats}
\usepackage{url}
\usepackage{verbatim}
\usepackage{graphicx}
\usepackage{cite}
\usepackage{booktabs}
\usepackage{multirow}
\usepackage[table]{xcolor}
\definecolor{lightblue}{RGB}{230,242,255}
\begin{document}
 
\title{SMAC: Spatial–Modal Joint Modeling and Adaptive Representation Collapse for Multimodal Object Tracking}

\author{
Gao Meijing,
Qitai Sun,
Huanyu Sun,
Bingxuan Yang,
Bingzhou Sun,
Xu Chen,
Yonghao Yan,
and Yuxuan Yang

\thanks{
Gao Meijing are with the School of Integrated Circuits and Electronics,
Beijing Institute of Technology, Beijing 100081, China, and with the
State Key Laboratory of Environment Characteristics and Effects for Near-space, China.

Qitai Sun, Huanyu Sun, Bingxuan Yang, Bingzhou Sun, Xu Chen, Yonghao Yan, and Yuxuan Yang
are with the School of Integrated Circuits and Electronics,
Beijing Institute of Technology, Beijing 100081, China.
}

\thanks{
This work was supported by the National Natural Science Foundation of China
under Grant 62471034 and the Sichuan Science and Technology Program
under Grant 2025NSFSC2078.
}

\thanks{
Corresponding author: Meijing Gao
(e-mail: gaomeijing@126.com).
}
}

\markboth{SMAC: Spatial–Modal Joint Modeling and Adaptive Representation Collapse}
{Gao Meijing et al.: SMAC}

\maketitle

\begin{abstract}
Multimodal multi-object tracking (MOT) under complex illumination remains challenging due to insufficient joint modeling of spatial and modal features and the limited adaptability of fixed fusion strategies. To address these issues, this paper proposes a spatial-modal convolution fusion and distillation-prompt-based multimodal MOT framework. A spatial-modal fusion backbone is first constructed, where a Basic module performs spatial feature extraction and modal interaction via decoupled 3D convolution, while a Mixed module models nonlinear cross-modal correlations through amplitude-phase decomposition. In addition, a representation collapse network is designed for adaptive multimodal fusion. A Distillation Prompt Guidance (DPG) module generates dynamic modal weights under teacher supervision, and a Global Modal Difference Aggregation (GMDA) module preserves discriminative information during multimodal representation collapse. Extensive experiments on the UniRTL dataset demonstrate the effectiveness of the proposed method. The proposed tracker achieves 63.31 HOTA and 79.21 MOTA on the RNT modality, outperforming several state-of-the-art methods while maintaining favorable inference efficiency. The source code and pretrained models are publicly available at https://github.com/QitaiSun/SMAC.
\end{abstract}

\begin{IEEEkeywords}
Multi-object tracking, multimodal fusion, RGB-NIR-TIR (RNT), spatial-modal convolution. 
\end{IEEEkeywords}

\section{Introduction}

\IEEEPARstart{M}{ulti}-Object Tracking (MOT) aims to continuously detect and associate multiple targets in video sequences, and has significant application value in intelligent surveillance, autonomous driving, and unmanned security systems. In recent years, deep learning-based tracking methods have achieved remarkable progress, with representative approaches including FairMOT~\cite{ref1}, ByteTrack~\cite{ref2}, and MOTRv2~\cite{ref3}. However, most existing methods are built upon a single RGB modality and are therefore susceptible to illumination degradation under low-light, over-exposure, and severe occlusion scenarios, resulting in weakened target representation capability and unstable identity association.

To improve robustness under complex environments, RGB-NIR-TIR multimodal perception has recently been introduced into MOT tasks. Among them, the NIR modality can provide relatively stable texture information under low-light conditions, while the TIR modality is insensitive to illumination variations and can effectively highlight thermal radiation regions of targets. The complementarity among multiple modalities provides a new solution for target detection and identity association in challenging scenarios. Nevertheless, existing multimodal MOT methods still suffer from the following limitations:

\begin{itemize}
    \item Existing methods rely on 2D convolution, concatenation, or attention-based weighting for cross-modal fusion, where spatial modeling and modality interaction are typically decoupled, limiting joint spatial–modal representation;
    
    \item Most approaches use fixed or weakly adaptive fusion strategies, making them sensitive to illumination variations and leading to unstable modality contributions in complex environments;
    
    \item Representation collapse from multimodal features to compact embeddings may result in loss of discriminative cross-modal information, affecting long-term identity consistency.
\end{itemize}

\begin{figure}[!t]
\centering
\includegraphics[width=3.4in]{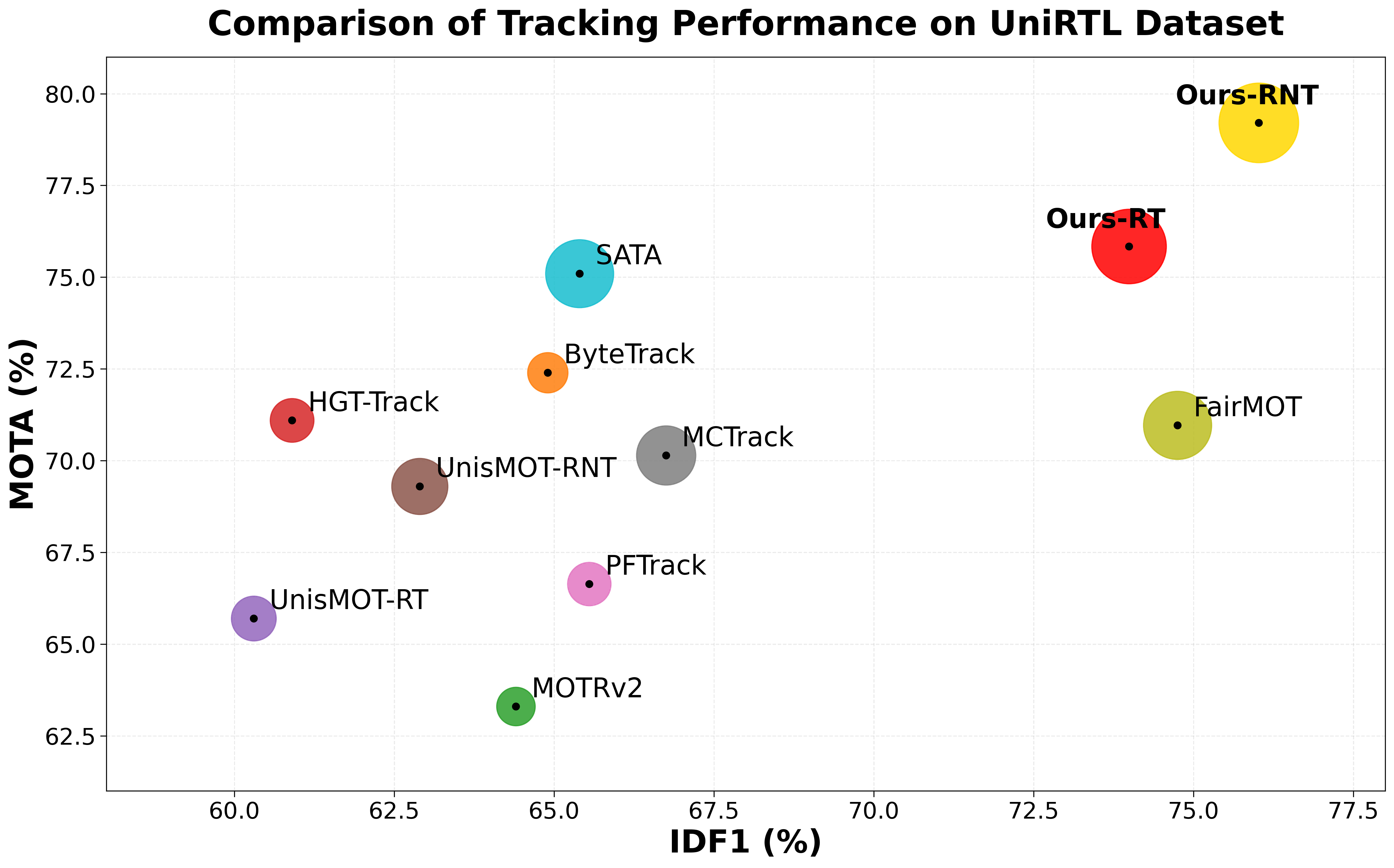}
\caption{Performance comparison of different MOT methods on UniRTL dataset.}
\label{fig1}

\vspace{1mm}
\begin{minipage}{0.95\linewidth}
\footnotesize
\textbf{Note:} The horizontal and vertical axes represent IDF1 and MOTA, respectively. The bubble size is proportional to the HOTA score, where larger bubbles indicate better overall tracking performance.
\end{minipage}

\end{figure}

To intuitively illustrate the differences among existing methods in terms of tracking accuracy, model complexity, and inference efficiency, Fig.~\ref{fig1} presents a performance comparison of several representative multimodal tracking methods. As shown in Fig.~\ref{fig1}, existing multimodal MOT approaches still struggle to achieve a satisfactory balance among tracking accuracy, model complexity, and inference efficiency. In contrast, the proposed SMAC achieves a superior accuracy--efficiency trade-off under challenging illumination conditions, demonstrating stronger overall performance advantages.

\begin{figure}[!t]
\centering
\includegraphics[width=3.4in]{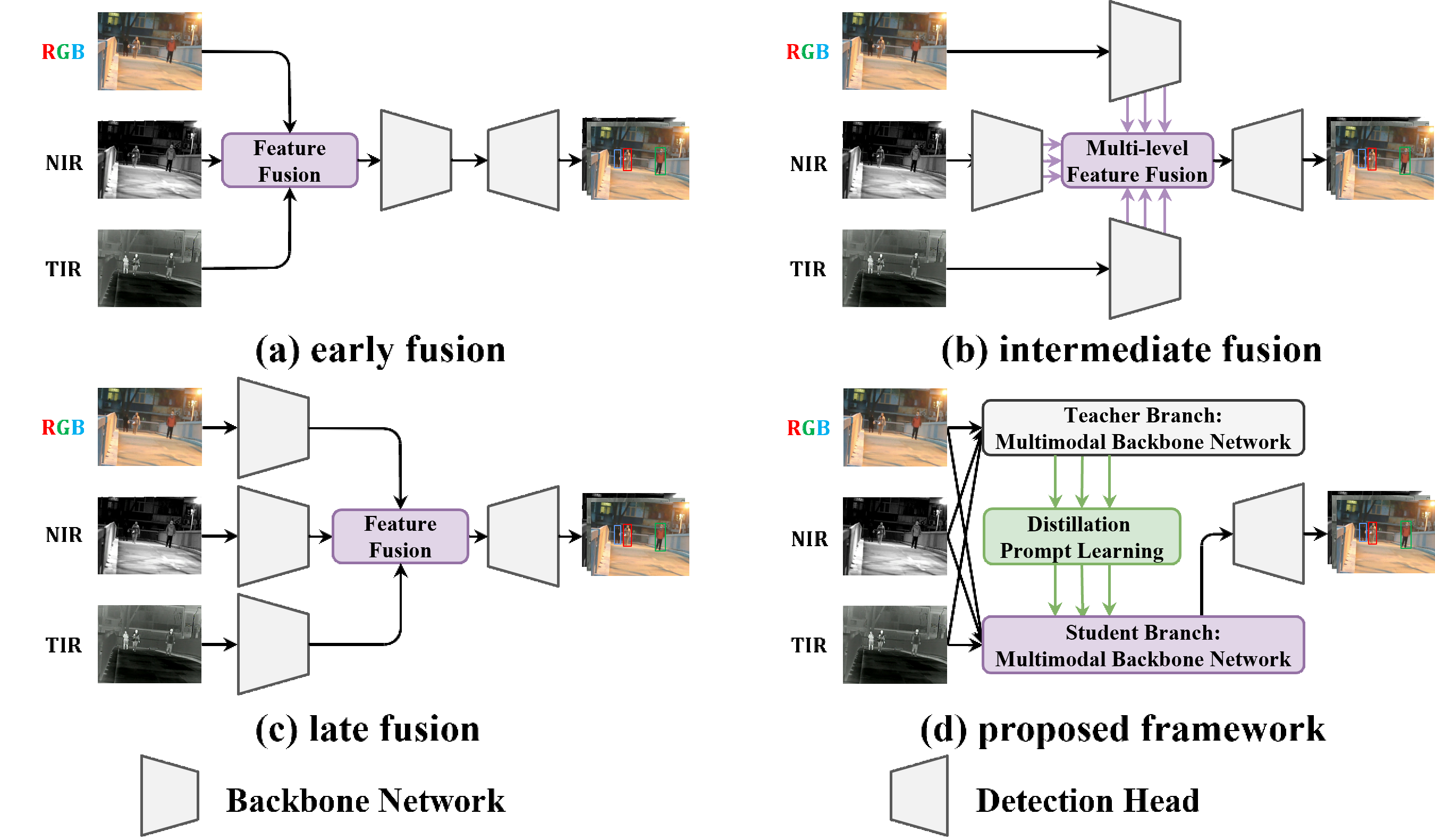}
\caption{Illustration of different multimodal feature fusion strategies, including early fusion, intermediate fusion, late fusion, and the proposed framework.}
\label{fig2}
\end{figure}

To address the above issues, this paper proposes a multimodal multi-object tracking framework based on spatial-modal convolution fusion and distillation prompts, termed SMAC. Different from traditional early fusion, intermediate fusion, and late fusion strategies, the proposed method constructs a unified multimodal backbone for RGB-NIR-TIR inputs, as illustrated in Fig.~\ref{fig2}. Specifically, a Basic module is designed in shallow layers to perform spatial feature extraction and modality interaction via decoupled 3D convolutions. In deeper layers, a Mixed module is introduced, which incorporates an amplitude--phase decomposition mechanism for modeling complex cross-modal correlations. Furthermore, a representation collapse network based on distillation prompts and global aggregation is proposed. The Distillation Prompt Guidance (DPG) module dynamically generates modality weights under teacher supervision for adaptive fusion under varying illumination conditions. Meanwhile, the Global Modal Difference Aggregation (GMDA) module enhances cross-modal discriminative representations by modeling modality commonality and discrepancy.

Extensive experiments are conducted on the UniRTL dataset~\cite{ref4}. Experimental results demonstrate that the proposed method achieves a favorable balance between tracking accuracy and inference efficiency. Under the RNT modality setting, the proposed tracker achieves 63.31 HOTA and 79.21 MOTA, outperforming existing state-of-the-art methods.

The main contributions of this paper are summarized as follows:

\begin{itemize}
    \item A spatial-modal convolution fusion framework for multimodal tracking is proposed, which achieves joint optimization of spatial and modal information through decoupled 3D convolutions and amplitude--phase modeling;
    
    \item A Distillation Prompt Guidance network and a Global Modal Difference Aggregation module are proposed to realize dynamic modality fusion and discriminative information preservation under complex illumination conditions;
    
    \item A unified multimodal multi-object tracking framework for RGB-NIR-TIR scenarios is constructed, effectively improving detection and association performance in complex environments;
    
    \item Extensive experiments on the UniRTL dataset verify the effectiveness of the proposed method in terms of both tracking accuracy and inference efficiency.
\end{itemize}

\section{Related Work}

This section reviews related studies in multi-object tracking. Section II-A introduces representative single-modal tracking paradigms, including TBD, SDE, JDE, and E2E methods. Section II-B summarizes recent advances in multimodal tracking and cross-modal representation learning.

\subsection{Single-Modal Multi-Object Tracking}

Existing single-modal MOT methods can generally be categorized into four paradigms: Tracking-by-Detection (TBD), Separate Detection and Embedding (SDE), Joint Detection and Embedding (JDE), and End-to-End Tracking (E2E).

TBD methods perform object detection followed by data association. Representative approaches include SORT~\cite{ref5}, Tracktor~\cite{ref6}, ByteTrack~\cite{ref2}, and OC-SORT~\cite{ref7}. Although efficient, they remain sensitive to detector quality and severe occlusion.

SDE methods introduce independent ReID networks to enhance identity preservation capability. Yu \emph{et al.}~\cite{ref8} verified the effectiveness of high-quality ReID features. Representative approaches include DeepSORT~\cite{ref9}, QDTrack~\cite{ref10}, BoT-SORT~\cite{ref11}, StrongSORT~\cite{ref12}, and Hybrid-SORT~\cite{ref13}, which improve association robustness through appearance modeling and motion compensation. However, the additional ReID branch introduces considerable computational overhead and remains sensitive to illumination degradation.

JDE methods jointly optimize detection and identity embedding within a shared framework, including JDE~\cite{ref14}, CenterTrack~\cite{ref15}, FairMOT~\cite{ref1}, CSTrack~\cite{ref16}, and RelationTrack~\cite{ref17}. Nevertheless, balancing detection and association objectives remains challenging.

E2E methods unify detection and association in a single framework, such as TransTrack~\cite{ref18}, TransMOT~\cite{ref19}, TrackFormer~\cite{ref20}, MOTR~\cite{ref21}, MOTRv2~\cite{ref3}, MeMOTR~\cite{ref22}, CO-MOT~\cite{ref23}, and MOTIP~\cite{ref24}. Despite their effectiveness, they generally require substantial computational resources.

\subsection{Multimodal Multi-Object Tracking}

Recent multimodal MOT methods mainly focus on RGB-T fusion and cross-modal association, including MTMMC~\cite{ref25}, HGT-Track~\cite{ref26}, UniRTL and UnisMOT~\cite{ref27}, VT-MOT and PFTrack~\cite{ref28}, and MCTrack~\cite{ref29}. Although these methods improve robustness under adverse conditions, most rely on attention-based or linear fusion mechanisms and lack explicit modeling of modality discrepancy and structural consistency.

Related multimodal fusion studies in 3D perception, such as MDRN~\cite{ref30}, PLPFusion~\cite{ref31}, PVF-DectNet++~\cite{ref32}, and SD$^2$-ReID~\cite{ref33}, improve representation learning through modality decoupling, consistency modeling, and adaptive aggregation. However, these methods are primarily designed for static-scene perception and do not explicitly address dynamic modality variation in tracking scenarios.

In summary, existing multimodal MOT methods still exhibit limitations in dynamic modality adaptation, cross-modal structural modeling, and unified multimodal representation learning under complex environments.

\section{Methodology} 

This section details the proposed method. The overall architecture of SMAC is first introduced in Section III-A. The design of the spatial-modal convolutional fusion backbone and the representation collapse network with distillation prompts and global aggregation are then described in Sections III-B and III-C, respectively.

\subsection{Overall Architecture}

Existing multimodal multi-object tracking methods mainly rely on static feature interaction or attention-based fusion mechanisms, which are insufficient for jointly modeling spatial structural consistency and dynamic modality variations under complex illumination conditions. As a result, they often suffer from modality interference, structural misalignment, and unstable feature representation in challenging scenarios such as low illumination, occlusion, and fast motion.

\begin{figure*}[t]
\centering
\includegraphics[width=\textwidth]{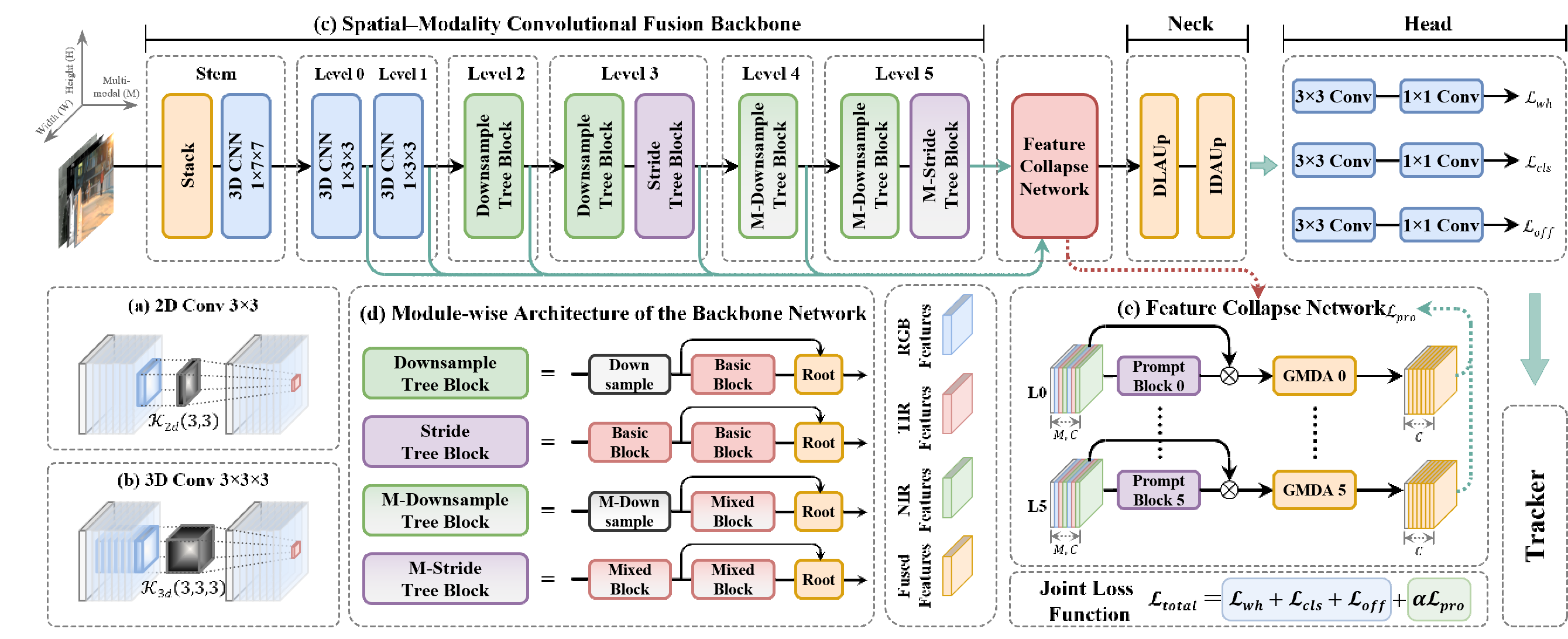}
\caption{Block diagram of the proposed multimodal object tracking framework based on spatial-modal convolutional fusion and distillation prompts.}
\label{fig:framework}
\end{figure*}

To address these issues, we propose a multimodal object tracking framework termed SMAC, which progressively enhances multimodal representation from three hierarchical levels: spatial–modal coupled modeling, dynamic modality recalibration, and structural representation collapse, as illustrated in Fig.~\ref{fig:framework}.

The proposed framework takes RGB, NIR, and TIR images as inputs, where each modality is represented as $\mathbb{R}^{3\times H\times W}$. As illustrated in Fig.~\ref{fig:framework}(a), conventional 2D convolution only performs feature extraction along spatial dimensions and cannot explicitly model cross-modal interactions. To jointly capture spatial structures and modality relationships within a unified framework, the convolution operation is further extended into a 3D representation space, as shown in Fig.~\ref{fig:framework}(b), enabling simultaneous modeling along both spatial and modality dimensions.

The overall framework mainly consists of two components. The first component is a spatial-modal convolutional fusion backbone for hierarchical feature extraction, whose overall architecture and internal module structures are illustrated in Fig.~\ref{fig:framework}(c) and Fig.~\ref{fig:framework}(d), respectively. Specifically, the framework first performs spatial-modal coupled modeling to explicitly capture spatial structural information and cross-modal interactions within a unified 3D feature space. Subsequently, a dynamic modality recalibration mechanism is introduced to adaptively estimate modality importance under varying illumination conditions, thereby alleviating modality imbalance and suppressing degraded modality responses.

The second component is a representation collapse network based on distillation prompts and global aggregation, as shown in Fig.~\ref{fig:framework}(e), which aims to collapse 3D multimodal representations into compact yet discriminative 2D embeddings while preserving cross-modal complementary structures and semantic consistency. The resulting 2D features are subsequently fed into the neck network for multi-scale feature aggregation, followed by the detection head for object classification and bounding box regression. Finally, during inference, a lightweight tracker based on IoU matching is employed to maintain identity consistency across video sequences.

\subsection{Spatial-Modal Convolutional Fusion Backbone}

\subsubsection{Backbone Overview}

\begin{figure}[t]
\centering
\includegraphics[width=\columnwidth]{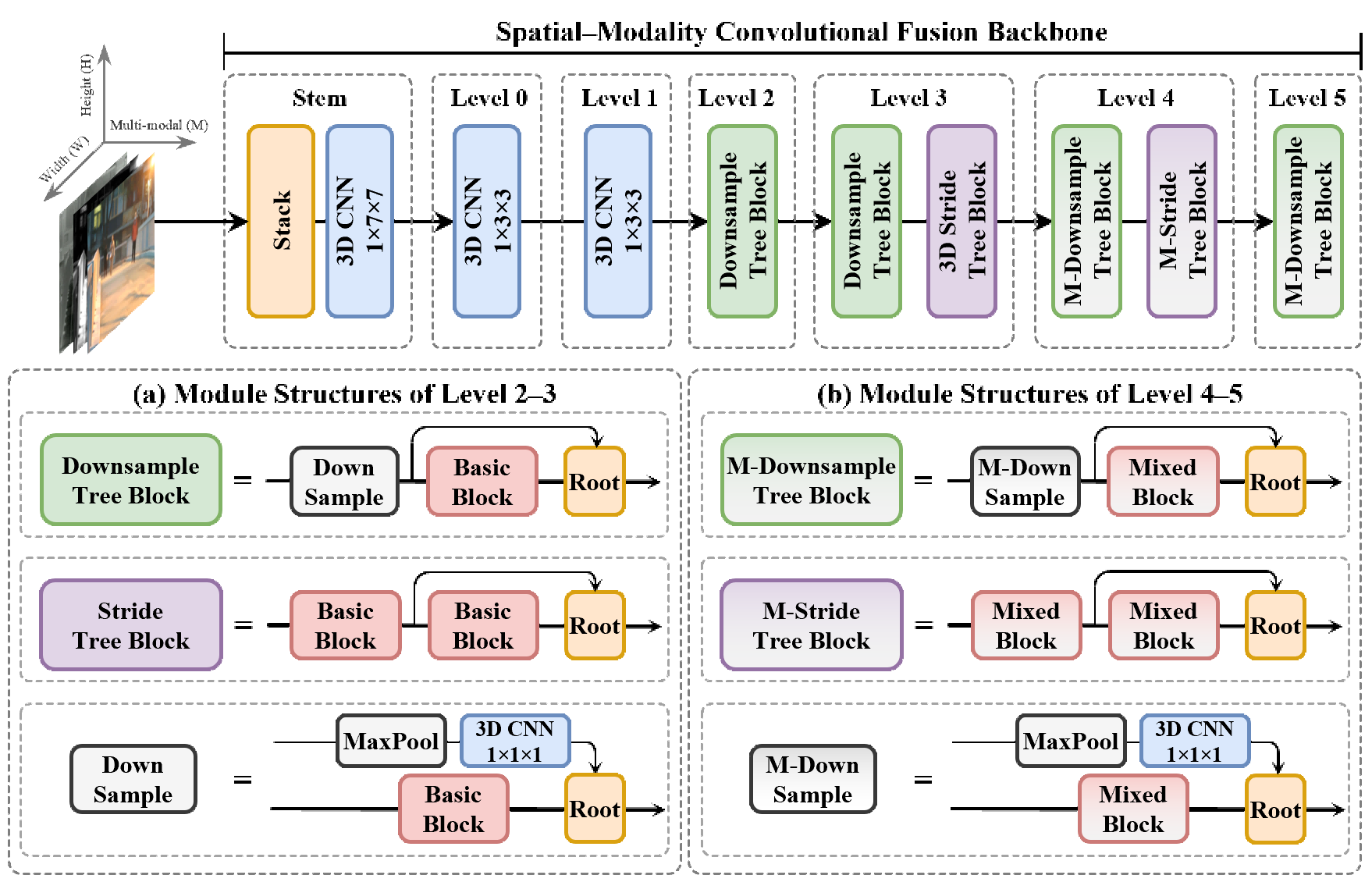}
\caption{Architecture of the spatial-modal convolutional fusion backbone.}
\label{fig:backbone}
\end{figure}

The architecture of the proposed spatial-modal convolutional fusion backbone is illustrated in Fig.~\ref{fig:backbone}.

The network consists of a Stem layer and multiple progressive feature extraction stages, following a hierarchical design principle from low-level spatial details to high-level semantic representations, thereby gradually enhancing multimodal feature expressiveness.

At shallow stages (Level 0--1), the network focuses on local spatial textures and edge structures, providing fundamental representations for subsequent cross-modal modeling. At intermediate stages (Level 2--3), a Basic module is introduced to decouple spatial and modality-specific information, alleviating the over-coupling issue commonly observed in standard 3D convolutions. At deep stages (Level 4--5), a Mixed module is further designed to enhance nonlinear cross-modal interactions in high-level semantic space, improving representation capability in complex scenarios.

The RGB, NIR, and TIR modalities are first stacked into a unified 3D tensor and processed by the Stem layer for initial feature extraction. The resulting features are then progressively fed into hierarchical modules for joint spatial-modal learning.

To address the high computational cost and feature entanglement issue of standard 3D convolutions, two low-rank decomposition-based designs are introduced: the Basic module and the Mixed module. The Basic module is used in intermediate layers to achieve lightweight decoupled modeling, reducing computational complexity while preserving representation capacity. In contrast, the Mixed module is employed in deep layers to strengthen nonlinear modeling ability and improve cross-modal interaction in high-level semantic representations.

Specifically, Levels 2--3 consist of Downsample Tree Blocks and Stride Tree Blocks, both adopting the Basic module for feature extraction. In Levels 4--5, the Basic module is replaced with the Mixed module, enabling stronger cross-modal interaction and fusion at higher semantic levels.

\subsubsection{Basic Module Design}

In the intermediate stages of the network (Levels 2--3), feature maps still retain relatively high spatial resolution and mainly contain low-level structural information such as edges and textures. Directly applying standard 3D convolutions for joint spatial-modal modeling at this stage introduces high computational complexity and may lead to premature entanglement between spatial representations and modality information, thereby weakening high-level semantic modeling capability in later stages.

To address this issue, this paper proposes a Basic module based on decoupled pseudo-3D convolution, which decomposes the conventional 3D convolution into two independent processes: spatial modeling and modality interaction. This design reduces computational cost while enabling effective cross-modal feature fusion.

\begin{figure}[t]
\centering
\includegraphics[width=\columnwidth]{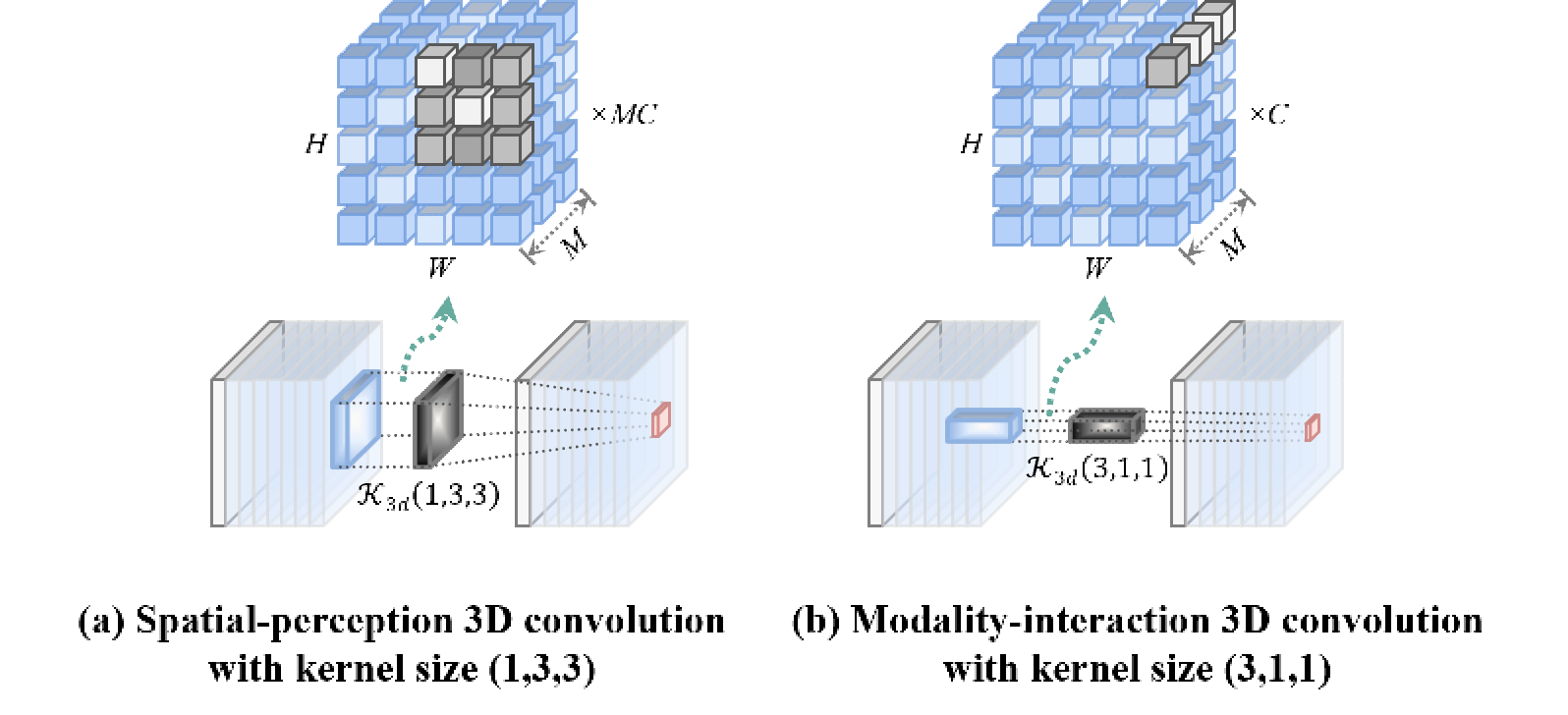}
\caption{3D convolution for modeling spatial and modality interaction.}
\label{fig:basic_3d}
\end{figure}

As shown in Fig.~\ref{fig:basic_3d}, the module first applies a spatial convolution with a kernel size of $1 \times 3 \times 3$ to independently extract local spatial structural features within each modality, thereby enhancing low-level texture and edge representations. Subsequently, a modality convolution with a kernel size of $3 \times 1 \times 1$ is introduced to enable cross-modal information interaction while preserving spatial consistency, thereby establishing explicit correlations among different modalities.

\begin{figure}[t]
\centering
\includegraphics[width=\columnwidth]{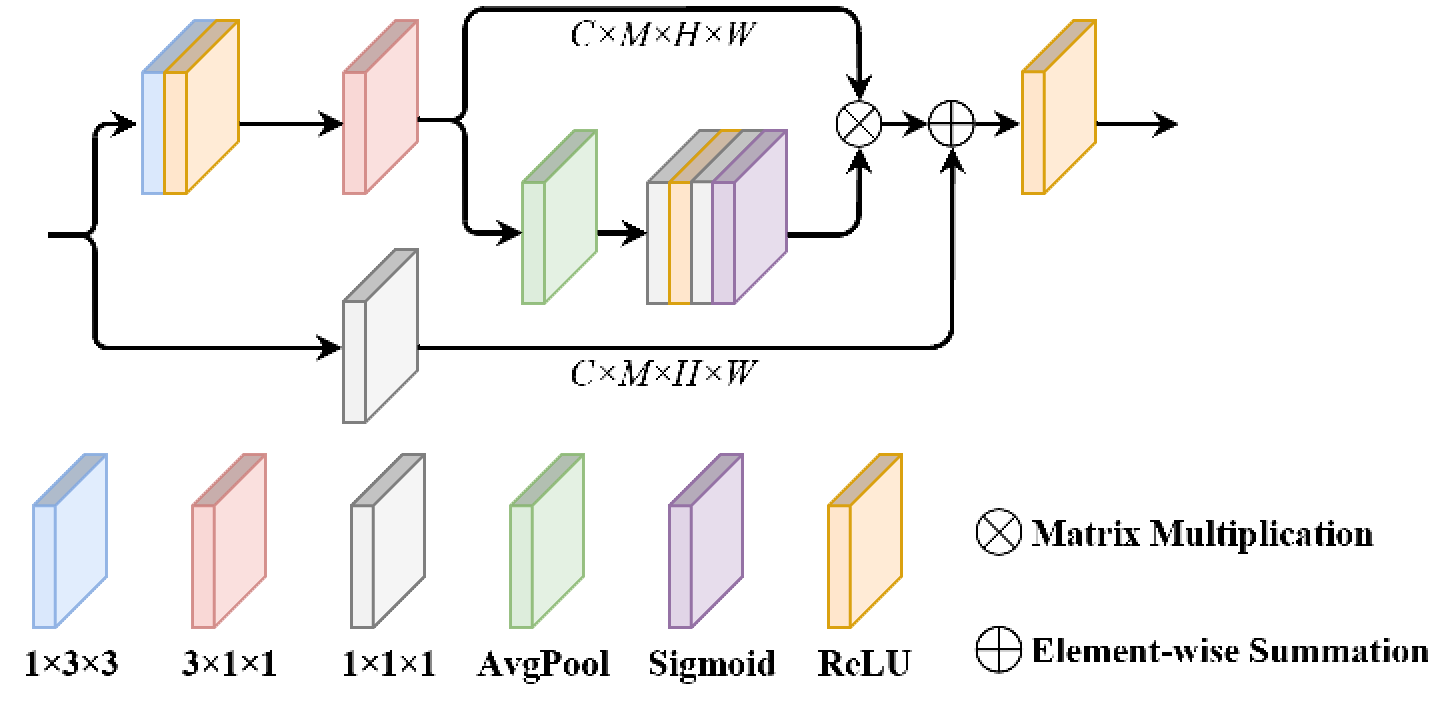}
\caption{Network architecture of the Basic module.}
\label{fig:basic_module}
\end{figure}

On this basis, as illustrated in Fig.~\ref{fig:basic_module}, the overall Basic module follows a pipeline of ``spatial decoupling -- modality interaction -- attention enhancement -- residual fusion.'' Specifically, the input multimodal features are first processed by grouped 3D convolutions for spatial feature extraction, followed by modality interaction convolution for cross-modal fusion. An MS-SE attention mechanism is then introduced to adaptively recalibrate the fused features, and finally a residual connection is applied to produce the enhanced feature representation.

Let the input multimodal feature be $\mathrm{X} \in \mathbb{R}^{C \times M \times H \times W}$, where $M$ denotes the number of modalities, and $C$, $H$, and $W$ denote the channel number, height, and width of the feature map, respectively.

Let the weight matrices of spatial decoupled convolution, modality interaction convolution, and cross-domain residual compensation be denoted as $\mathrm{W}_s \in \mathbb{R}^{C \times 1 \times 3 \times 3}$, $\mathbf{W}_m \in \mathbb{R}^{C \times 3 \times 1 \times 1}$, $\mathrm{W}_r \in \mathbb{R}^{C \times 1 \times 1 \times 1}$,
respectively.

First, spatial feature extraction is formulated as:
\begin{equation}
\mathrm{X}_s = \delta\left(\mathcal{B}\left(\mathrm{W}_s \odot_g \mathrm{X}\right)\right)
\end{equation}

Then, cross-modal interaction is performed based on the extracted spatial features:
\begin{equation}
\mathrm{X}_m = \mathcal{B}\left(\mathrm{W}_m \odot \mathrm{X}_s\right)
\end{equation}

Finally, the interacted features are fed into the attention module for adaptive refinement, and aggregated with the cross-domain residual compensation branch to produce the output:
\begin{equation}
\mathrm{X}_{out} =
\delta\left(
f_{\mathrm{MS\text{-}SE}}(\mathrm{X}_m)
+
\mathcal{B}\left(\mathrm{W}_r \odot \mathrm{X}\right)
\right)
\end{equation}

Here, $\odot_g$ denotes the grouped 3D convolution operation based on a depthwise separable design, $\odot$ denotes standard 3D convolution, $\delta(\cdot)$ denotes the ReLU activation function, $f_{\mathrm{MS\text{-}SE}}(\cdot)$ denotes the attention enhancement mapping function, and $\mathcal{B}(\cdot)$ denotes the residual compensation branch consisting of 3D batch normalization (BN3D) and activation.

Compared with standard 3D convolution, the Basic module explicitly decouples spatial and modality modeling, effectively reducing both parameter count and computational complexity. Moreover, the grouped convolution strategy further reduces redundant computation, enabling higher inference efficiency while preserving representational capacity. The design of this module aims to maintain complete spatial structure modeling while preventing premature modality entanglement, thereby preserving independent modality representation spaces for subsequent high-level semantic modeling.

\subsubsection{Mixed Module Design}

As the network depth increases (Levels 4--5), multimodal features gradually enter a high-level semantic space, where cross-modal correlations are reflected not only in response intensity differences but also in the consistency of target structural information and spatial topological relationships. Traditional linear convolutions or attention-based fusion methods mainly focus on response magnitude while lacking explicit modeling of cross-modal structural consistency. Consequently, they are prone to structural misalignment and modality interference under challenging conditions such as complex illumination, occlusion, and modality degradation.

From the perspective of frequency-domain representation, amplitude mainly reflects feature response intensity and semantic saliency, whereas phase primarily determines spatial structural distribution and edge topology. For multimodal inputs such as RGB, NIR, and TIR, although significant differences exist in texture and response intensity across modalities, the target regions usually share highly consistent spatial structures. Therefore, compared with direct linear feature fusion, utilizing phase information to model cross-modal structural consistency can more effectively alleviate response deviation among modalities.

\begin{figure}[t]
\centering
\includegraphics[width=\columnwidth]{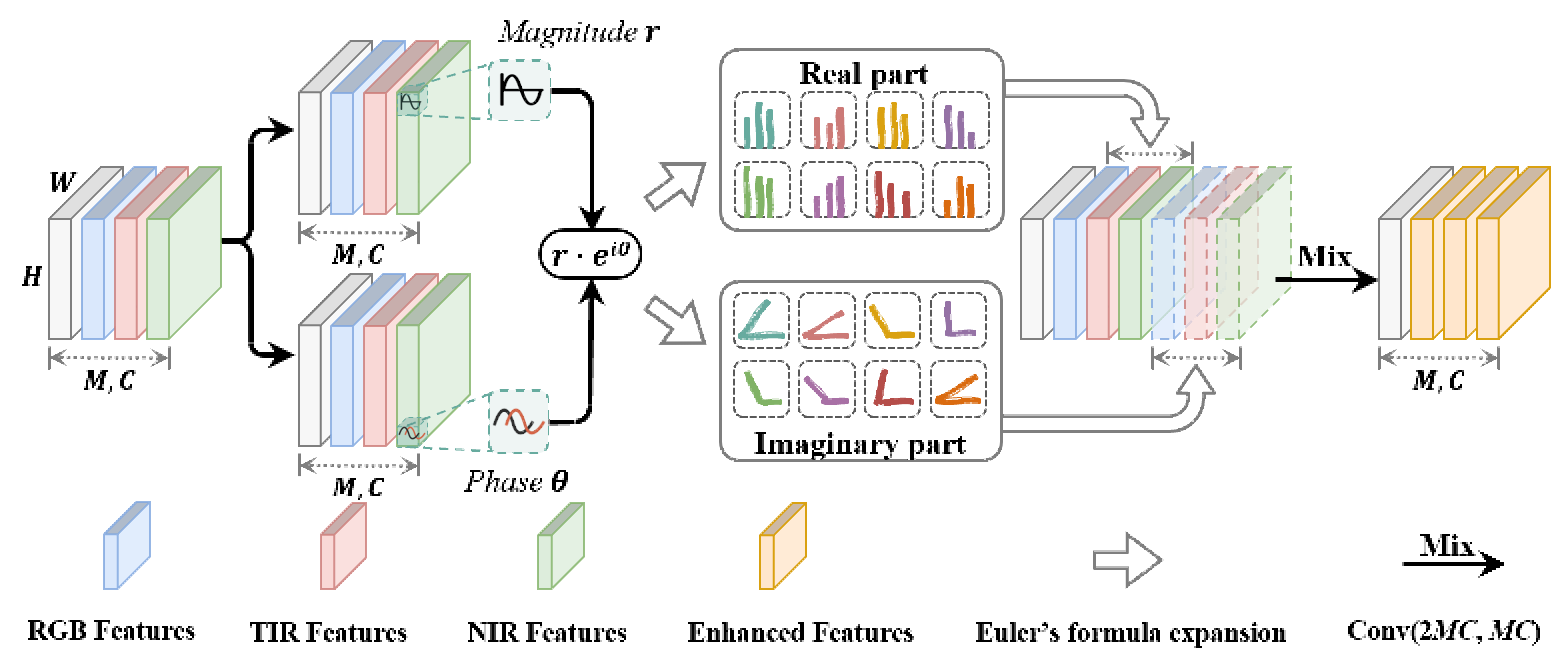}
\caption{Illustration of the Mixed convolution structure based on amplitude--phase modeling.}
\label{fig:mixed_structure}
\end{figure}
To this end, this paper designs a Mixed module with stronger representational capability by introducing an amplitude--phase decomposition mechanism to decouple modality response intensity and structural relationships. Specifically, the amplitude component is used to describe semantic responses and saliency information, while the phase component is employed to preserve cross-modal structural consistency and spatial dependency relationships. In addition, complex-valued feature representations are introduced to jointly model complementary and differential multimodal information.

As shown in Fig.~\ref{fig:mixed_structure}, the overall structure of the module remains consistent with the Basic module, where spatially decoupled convolution is still adopted in the front-end to preserve spatial structure modeling capability, formulated as:
\begin{equation}
\mathrm{X'}_s = \delta\left(\mathcal{B}\left(\mathrm{W}_s \odot_g \mathrm{X}\right)\right)
\end{equation}

Different from the Basic module, during the modality interaction stage, the Mixed module introduces the MS-Mixed convolution to replace the conventional linear mapping operator. Through a more sophisticated modality mixing mechanism, nonlinear feature transformation can be achieved, which is formulated as:
\begin{equation}
\mathrm{X}_{out} =
\delta\left(
\mathcal{F}_{mix}(\mathrm{X'}_s)
+
\mathcal{B}\left(\mathrm{W}_r \odot \mathrm{X}\right)
\right)
\end{equation}

In the Mixed module, inspired by the idea of Wave-MLP~\cite{ref34}, an amplitude--phase decomposition mechanism is introduced for structured feature modeling. Let the input multimodal feature be represented as 
$M=\left[m_1,m_2,\cdots,m_N\right]\in\mathbb{R}^{C\times H\times W}$, 
where $N$ denotes the number of modalities, $m_i$ represents the feature map of the $i$-th modality, and $C$, $H$, and $W$ denote the channel number, height, and width of the feature map, respectively. Specifically, the amplitude component is utilized to characterize feature response intensity and contextual information, while the phase component is employed to describe structural discrepancies and relational information across modalities. The formulation is given as follows:

\begin{equation}
|\mathrm{r}_i| = W^m \odot m_i, \quad i = 1,2,\cdots,N
\end{equation}

\begin{equation}
\theta_i = \delta\left(W^p \odot m_i\right), \quad i = 1,2,\cdots,N
\end{equation}

where $\mathrm{W}^m$ and $\mathrm{W}^p$ denote the learnable parameters for amplitude modeling and phase modeling, respectively; $\delta(\cdot)$ denotes the ReLU activation function; and $|\mathrm{r}_i|$ and $\theta_i$ represent the amplitude and phase components of the $i$-th modality, respectively.

Furthermore, the amplitude and phase components are combined through Euler’s formula to construct complex-valued feature representations:

\begin{equation}
\widetilde{\mathrm{m}}_i =
|\mathrm{r}_i| \odot \cos(\theta_i)
+ i |\mathrm{r}_i| \odot \sin(\theta_i),
\quad i = 1,2,\cdots,N
\end{equation}

where the real part $|\mathrm{r}_i|\cos\theta_i$ is used to encode modality-shared structural semantic information, while the imaginary part $|\mathrm{r}_i|\sin\theta_i$ is utilized to model modality-specific differential information, thereby enabling complementary representation learning.

Based on this formulation, complex-valued features are further aggregated through spatial-modal convolution. The output of the $i$-th modality is expressed as:

\begin{equation}
\widetilde{\mathrm{o}}_i =
\sum_{n \in N(i)} K(n) \odot \widetilde{\mathrm{m}}_n + \widetilde{\mathrm{m}}_i,
\quad i = 1,2,\cdots,N
\end{equation}

where $N(i)$ denotes the feature set within the local spatial-modal neighborhood of the $i$-th modality feature, $K(n)$ denotes the convolution kernel weight corresponding to each neighboring position, and $\mathrm{m}_i$ represents the residual term introduced to enhance training stability.

Finally, the real and imaginary components are fused to obtain the real-valued output feature $\mathrm{o}_i$:

\begin{equation}
\begin{aligned}
\mathrm{o}_i
=&
\sum_{n \in N(i)} K(n)
\Big(
|\mathrm{r}_n| \cos\theta_n \\
&
\quad +
|\mathrm{r}_n| \sin\theta_n
\Big)
+ m_i, \\
&
\quad i = 1,2,\cdots,N
\end{aligned}
\end{equation}

As indicated by Eq.~(10), the proposed mechanism introduces dynamic modulation weights during neighborhood aggregation, enabling adaptive weighted fusion of different modalities and their local spatial neighborhood features. Consequently, the model achieves stronger joint modeling capability for both cross-modal discrepancy and complementarity.

\begin{figure}[t]
\centering
\includegraphics[width=\columnwidth]{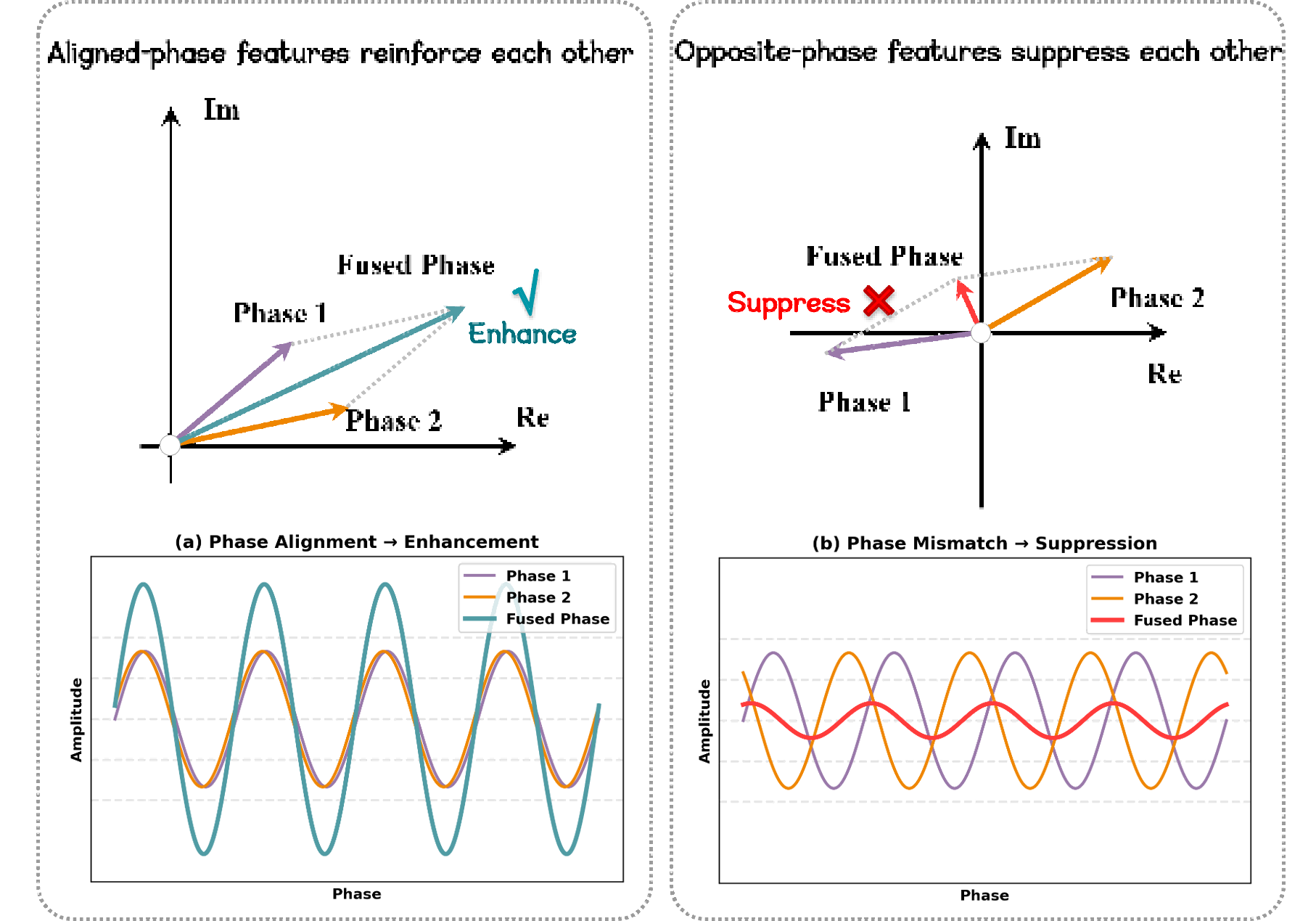}
\caption{Illustration of the cross-modal dynamic fusion mechanism based on amplitude--phase decomposition.}
\label{fig:dynamic_fusion}
\end{figure}

Therefore, as illustrated in Fig.~\ref{fig:dynamic_fusion}, the proposed module can simultaneously enhance the representation capability of modality-shared semantic information and modality-specific differential information, enabling more sufficient fusion and interaction of multimodal features within the deep semantic space.

In summary, through the amplitude--phase decomposition and complex-valued modeling mechanisms, the Mixed module further improves the representational capability and robustness of cross-modal relationship modeling while preserving spatial structural modeling ability.

\subsection{Adaptive Representation Collapse Network Based on Distillation Prompt and Global Aggregation}

To map three-dimensional multimodal features into compact and more discriminative two-dimensional representations, this paper constructs a representation collapse network based on distillation prompts and global aggregation. This module is designed to address two key challenges in multimodal fusion. First, the importance of different modalities varies dynamically across different scenarios, making traditional fixed fusion strategies insufficient for adaptive modeling. Second, during feature dimensionality reduction, cross-modal differential information is prone to information loss during aggregation, thereby weakening representation capability.

To address the above issues, a two-stage representation collapse mechanism composed of a Distillation Prompt Guidance (DPG) module and a Global Modal Difference Aggregation (GMDA) module is proposed. Specifically, the DPG module focuses on dynamic modality weighting at the sample level, while the GMDA module performs structured feature reconstruction at the representation level.

\subsubsection{Distillation Prompt-Guided Network (DPG)}

In multimodal scenarios, the contribution of different modalities to target representation exhibits significant scene dependency. To address this issue, this paper proposes a Distillation Prompt-Guided (DPG) network, which achieves adaptive learning of modality weights through a teacher--student paradigm.

\begin{figure}[t]
\centering
\includegraphics[width=\columnwidth]{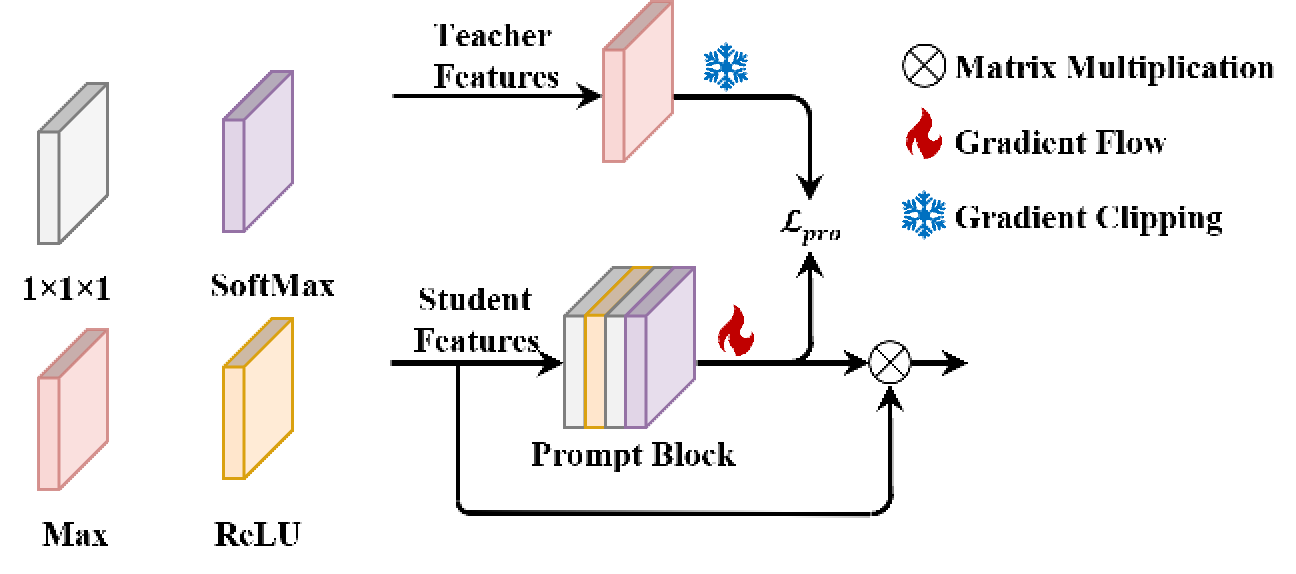}
\caption{Learning pipeline of the Distillation Prompt-Guided Network}
\label{fig:dpg}
\end{figure}

As illustrated in Fig.~\ref{fig:dpg}, the DPG adopts a teacher--student architecture. The teacher branch utilizes the maximum response of multimodal features as the supervision signal, while the student branch generates modality fusion weights through a lightweight prompt module and optimizes them via distillation loss, thereby achieving dynamic multimodal fusion.

The student branch first employs two convolutional layers to generate modality attention weights $W \in \mathbb{R}^{B \times 1 \times 3 \times H \times W}$:
\begin{equation}
A = \delta(\operatorname{W}_2 \odot \delta(\operatorname{W}_1 \odot X))
\end{equation}

\begin{equation}
W = \operatorname{SoftMax}_{dim=2}(A)
\end{equation}

where $X \in \mathbb{R}^{B \times C \times 3 \times H \times W}$ denotes the input multimodal features containing RGB, NIR, and TIR modalities, $\operatorname{W}_1$ and $\operatorname{W}_2$ represent the parameters of the two convolutional layers, $\odot$ denotes the 3D convolution operation, and $\delta(\cdot)$ denotes the ReLU activation function. Subsequently, Softmax normalization is performed along the modality dimension to obtain the cross-modal weight tensor.

Based on the learned weights, the multimodal features are adaptively fused to obtain the fused representation:
\begin{equation}
F_{low} = \sum_{i \in \{rgb,nir,tir\}} W_i \otimes F_i
\end{equation}

where $F_{low}$ denotes the fused feature representation, $W_i$ denotes the weight slice corresponding to the i-th modality, and $\otimes$ represents broadcast-wise element multiplication. The resulting output $F_{low} \in \mathbb{R}^{B \times C \times H \times W}$ corresponds to the weighted fusion result of multimodal features.

Different from the student branch, the teacher branch adopts a statistically driven robust representation strategy by constructing supervision signals through maximum responses along the modality dimension, thereby capturing the most discriminative modality activation:
\begin{equation}
F_{max} = \operatorname{Max}_{dim=2}(F_t)
\end{equation}

where $F_t \in \mathbb{R}^{B \times C \times 3 \times H \times W}$ denotes the multimodal features extracted by the teacher backbone network, and $F_{max}$ denotes the maximum-response activation.

The core idea of this design is that the maximum-response feature can reflect the most discriminative modality information under the current scene, thereby providing stable structural supervision for the student branch.

During training, the teacher network is frozen while only the student branch is optimized. Mean Squared Error (MSE) is adopted to construct the distillation loss function, enabling the student branch to learn dynamic modality importance distributions. Through this mechanism, the DPG achieves scene-aware adaptive modality prompt modeling capability.

\subsubsection{Global Modal Difference Aggregation (GMDA)}

To further improve information fidelity during multimodal representation collapse, this paper proposes a Global Modal Difference Aggregation (GMDA), which explicitly models cross-modal differential structures and achieves high-quality collapsing of 3D multimodal features into compact yet discriminative 2D representations.

\begin{figure}[!t]
\centering
\includegraphics[width=\columnwidth]{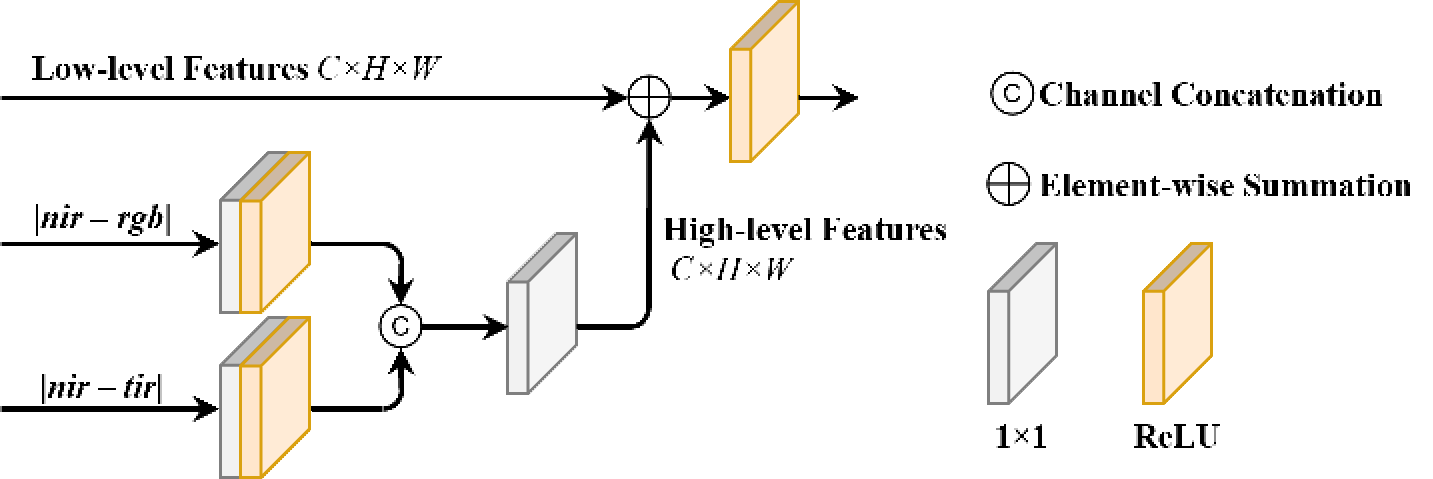}
\caption{Architecture of the GMDA module}
\label{fig:gmda}
\end{figure}

As illustrated in Fig.~\ref{fig:gmda}, the GMDA is built upon the output features of the DPG module and introduces a cross-modal difference modeling strategy. Considering that the NIR spectrum lies between RGB and TIR in the spectral domain and can simultaneously reflect the physical differences between visible and thermal infrared modalities, this paper employs NIR features as the reference modality to construct cross-modal differential representations:

\begin{equation}
\mathrm{\Delta}_{\mathrm{tn}} = \delta \left( \mathrm{\operatorname{W}}_1 * \left| \mathrm{\operatorname{F}}_{\mathrm{tir}} - \mathrm{\operatorname{F}}_{\mathrm{nir}} \right| \right)
\end{equation}

\begin{equation}
\mathrm{\Delta}_{\mathrm{rn}} = \delta \left( \mathrm{\operatorname{W}}_2 * \left| \mathrm{\operatorname{F}}_{\mathrm{rgb}} - \mathrm{\operatorname{F}}_{\mathrm{nir}} \right| \right)
\end{equation}

where $\mathrm{\operatorname{F}}_{\mathrm{rgb}}, \mathrm{\operatorname{F}}_{\mathrm{nir}}, \mathrm{\operatorname{F}}_{\mathrm{tir}} \in \mathbb{R}^{B \times C \times H \times W}$ denote the RGB, NIR, and TIR features extracted by the backbone network, respectively; $\mathrm{\operatorname{W}}_1$ and $\mathrm{\operatorname{W}}_2$ represent the weight matrices of $1 \times 1$ convolutions; $*$ denotes the 2D convolution operation; and $\delta(\cdot)$ denotes the ReLU nonlinear activation function.

Subsequently, the differential features are concatenated along the channel dimension and projected through a $1 \times 1$ convolution to generate the high-order feature representation $\mathrm{\operatorname{F}}_{\mathrm{high}}$:

\begin{equation}
\mathrm{\operatorname{F}}_{\mathrm{high}} = \mathcal{B} \left( \mathrm{\operatorname{W}}_{\mathrm{diff}} * \mathrm{Concat} \left( \mathrm{\Delta}_{\mathrm{tn}}, \mathrm{\Delta}_{\mathrm{rn}} \right) \right)
\end{equation}

where $\mathrm{\operatorname{W}}_{\mathrm{diff}}$ denotes the 2D convolution weight, and $\mathcal{B}(\cdot)$ denotes the 2D Batch Normalization operation.

Furthermore, the high-order feature $\mathrm{\operatorname{F}}_{\mathrm{high}}$ is utilized as a global modulation term and fused with the low-order common feature $\mathrm{\operatorname{F}}_{\mathrm{low}}$ through residual aggregation, thereby preserving structural information and enhancing semantic representation during representation collapse. The final fused 2D feature map $\mathrm{\operatorname{F}}_{\mathrm{out}}$ is obtained as:

\begin{equation}
\mathrm{\operatorname{F}}_{\mathrm{out}} = \delta \left( \mathrm{\operatorname{F}}_{\mathrm{low}} + \mathrm{\operatorname{F}}_{\mathrm{high}} \right)
\end{equation}

This process essentially constructs a ``difference-driven representation collapse mechanism,'' which explicitly preserves cross-modal conflicts and complementary structures during multimodal representation collapse, thereby enhancing the representational capability and discriminative power of the final 2D embeddings.

\section{Experiments}

This section presents the experimental results and analysis of the proposed method. Section IV-A introduces the dataset, experimental platform, training settings, and evaluation metrics. Section IV-B investigates the effectiveness of different convolution designs and network components through ablation studies. Section IV-C compares the proposed method with state-of-the-art approaches in terms of tracking performance, illumination robustness, and inference efficiency. Finally, Section IV-D provides qualitative visualizations to further analyze the proposed multimodal fusion mechanism and tracking behavior.

\subsection{Experimental Settings}

\subsubsection{Dataset and Experimental Platform}

Experiments are conducted on the UniRTL dataset. The dataset contains RGB, NIR, and TIR modalities, consisting of 50 multi-object tracking video sequences covering low-, medium-, and high-illumination scenarios, which are used to evaluate the robustness of multimodal object tracking under complex lighting conditions.

According to the official split protocol, the dataset is divided into training and validation sets for model training and performance evaluation. The dataset provides strict cross-modal spatial alignment for evaluating multimodal fusion methods.

All experiments are implemented based on the PyTorch and Detectron2 frameworks and conducted under a unified hardware environment to ensure fairness and reproducibility of results.

\subsubsection{Training and Evaluation Details}

During training, the SGD optimizer is adopted, combined with a Warmup + Cosine learning rate decay strategy. The initial learning rate is set to $1\times10^{-4}$, momentum is 0.9, and the total training epochs are 30. The batch size is set to 16 during training and 4 during testing, and the confidence threshold is set to 0.5.

To improve model robustness, consistent data augmentation strategies are applied to RGB, NIR, and TIR modalities, including random color jittering, scale transformations, and geometric transformations.

Performance evaluation follows standard multi-object tracking metrics (HOTA, DetA, AssA, MOTA, etc.), providing a comprehensive analysis from both detection and association perspectives.

\subsection{Ablation Study}

\subsubsection{Convolution Design Ablation in the Basic Module}

To verify the effectiveness of each component, ablation studies are conducted on the convolution design of the Basic module and key modules of the overall network. The results are shown in Table~\ref{tab:basic_ablation}.

\begin{table}[htbp]
\centering
\caption{Ablation results of the Basic module}
\label{tab:basic_ablation}
\footnotesize
\setlength{\tabcolsep}{4pt}

\begin{tabular}{cccccccc}
\toprule
Module & Conv & MS-SE & mAP$\uparrow$ & HOTA$\uparrow$ & DetA$\uparrow$ & AssA$\uparrow$ \\
\midrule
A & 2D-SS & w/o & 43.60 & 59.46 & 60.48 & 58.90 \\
B & 3D-SS & w/o & 47.03 & 60.48 & 62.75 & 58.90 \\
C & 3D-All & w/o & 45.66 & 60.05 & 61.50 & 59.18 \\
D & 3D-MS & w/o & \textbf{50.01} & 62.71 & \underline{64.92} & 61.13 \\
E & 3D-SM & w/o & 49.25 & \underline{62.74} & 64.57 & \underline{61.52} \\
F & 3D-SM & w/  & \underline{49.27} & \textbf{63.31} & \textbf{64.93} & \textbf{62.39} \\
\bottomrule
\end{tabular}

\vspace{2mm}

\begin{minipage}{0.95\linewidth}
\footnotesize
\textbf{Note:} ``w/o'' denotes without the MS-SE module, and ``w/'' denotes with the MS-SE module. A unified evaluation protocol is used across all experiments, where the best results are highlighted in \textbf{bold} and the second-best results are indicated by \underline{underlining}.
\end{minipage}

\end{table}

The experimental results demonstrate that, compared with 2D convolution and full 3D convolution structures, introducing a spatial–modal decoupled 3D convolution design consistently improves all evaluation metrics. This indicates that explicitly separating spatial and modality information enhances feature representation capability. Furthermore, introducing the MS-SE attention mechanism on top of the decoupled convolution further improves performance, confirming the effectiveness of attention-based feature recalibration for cross-modal enhancement.

Overall, spatial–modal decoupled convolution and the MS-SE attention mechanism exhibit strong complementarity, improving both detection performance and cross-modal representation ability. Therefore, Model F is selected as the default configuration of the Basic module.

\subsubsection{Component Contribution Analysis}

To systematically evaluate the contribution of each component to the overall performance, we progressively introduce 3D convolution (3D Conv), the Basic module, the Mixed module, the distillation prompt-guided network (DPG), and the global modal difference aggregation module (GMDA). The results are shown in Table~\ref{tab:ablation}.

\begin{table}[!t]
\centering
\caption{Ablation results of key components}
\label{tab:ablation}
\renewcommand{\arraystretch}{1.15}
\setlength{\tabcolsep}{5pt}

\begin{tabular}{ccccc}
\toprule
Variant & mAP$\uparrow$ & HOTA$\uparrow$ & DetA$\uparrow$ & AssA$\uparrow$ \\
\midrule
A & 43.60 & 59.46 & 60.48 & 58.90 \\
B & 47.03 & 60.48 & 62.75 & 58.90 \\
C & 48.72 & 61.95 & 63.97 & \underline{60.61} \\
D & \underline{49.75} & \underline{62.36} & \underline{64.82} & 60.53 \\
E & 47.92 & 61.43 & 63.20 & 60.32 \\
F & 46.53 & 61.00 & 62.66 & 59.95 \\
G & \textbf{49.81} & \textbf{63.56} & \textbf{65.81} & \textbf{61.89} \\
\bottomrule
\end{tabular}

\vspace{1.5mm}

\begin{minipage}{0.48\textwidth}
\footnotesize
\textbf{Note:} A: all modules are removed; B: only 3D Conv is used; C: 3D Conv + Basic; D: 3D Conv + Basic + Mixed; E: 3D Conv + DPG; F: 3D Conv + GMDA; G: all modules are used.
\end{minipage}

\end{table}

To further analyze the contribution of each component to tracking performance, we progressively introduce each module. The results demonstrate consistent improvements across all metrics.

The Basic and Mixed modules mainly enhance local spatial–modal representation and detection performance, while DPG and GMDA primarily improve dynamic cross-modal fusion and association modeling.

When all modules are jointly used, the model achieves the best performance across all metrics, demonstrating strong complementarity among components and significantly improving robustness and stability in complex scenarios.

\begin{figure}[!t]
\centering
\includegraphics[width=0.48\textwidth]{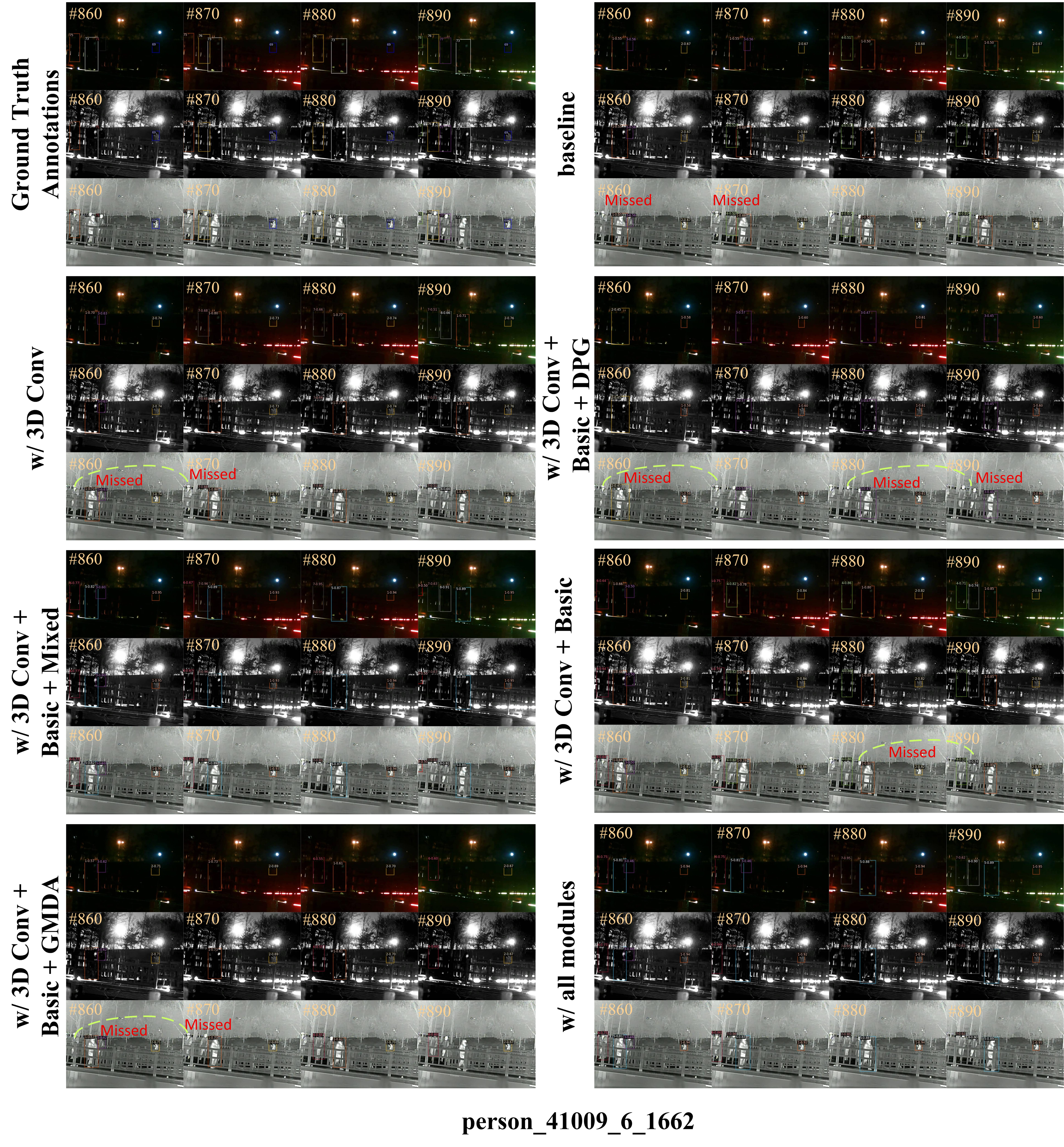}
\caption{Visualization comparison of different module combinations under crowded occlusion and fast-motion scenarios.}
\label{fig11}
\end{figure}

To further intuitively analyze the tracking performance under complex conditions, we select crowded, fast-moving, and heavily occluded scenarios for visualization, as shown in Fig.~\ref{fig11}. The baseline model suffers from missed detections, identity switches, and trajectory interruptions under these conditions.

As the Basic, Mixed, DPG, and GMDA modules are progressively introduced, tracking stability and identity consistency are gradually improved. Notably, the full model can still recover correct trajectories after long-term occlusion and maintain stable identity association, verifying the effectiveness of the proposed spatial–modal joint modeling and dynamic fusion strategy.

Overall, the Basic and Mixed modules contribute more to detection accuracy, while DPG and GMDA mainly improve association performance. The combination of all modules further unlocks full potential, achieving optimal performance in both detection and tracking tasks.

In summary, the proposed framework exhibits a clear functional decomposition: shallow spatial–modal decoupling enhances feature robustness, deep nonlinear fusion improves semantic representation, and adaptive cross-modal reweighting improves tracking stability under varying illumination conditions.

\subsection{Comparison With State-of-the-Arts}

\subsubsection{Overall Performance Comparison}

In this work, a lightweight tracking strategy based on Kalman filtering and IoU matching is adopted. Comprehensive metric comparisons are conducted with several state-of-the-art methods on the validation set, and the results are reported in Table~\ref{tab:sota_tracking}. Since most existing methods only support visible-light and thermal infrared dual-modal input (RT), experiments under this setting are first conducted. Only UnisMOT~\cite{ref27} supports tri-modal input (RNT); therefore, evaluations are also performed under the RNT setting. All compared methods are published in top-tier computer vision conferences or journals and are highly representative.

\begin{table*}[!t]
\centering
\caption{Comparison of tracking performance with existing state-of-the-art methods}
\label{tab:sota_tracking}
\renewcommand{\arraystretch}{1.15}

\resizebox{\textwidth}{!}{
\begin{tabular}{cccccccccc}
\toprule
Methods & Venue & Year & Backbone & Modal & HOTA$\uparrow$ & DetA$\uparrow$ & AssA$\uparrow$ & MOTA$\uparrow$ & IDF1$\uparrow$ \\
\midrule

UniTrack~\cite{ref35} & NeurIPS & 2021 & ResNet-50 & RT & 46.90 & 46.10 & 48.40 & 57.10 & 57.40 \\
Unicorn~\cite{ref36} & ECCV & 2022 & ResNet-50 & RT & 49.40 & 48.30 & 51.20 & 59.00 & 60.60 \\
ByteTrack~\cite{ref2} & ECCV & 2022 & CSPDarknet & RT & 53.40 & 58.20 & 49.60 & 72.40 & 64.90 \\
MOTRv2~\cite{ref3} & CVPR & 2023 & ResNet-50 & RT & 53.10 & 51.20 & 55.80 & 63.30 & 64.40 \\
HGT-Track~\cite{ref26} & TMM & 2024 & PVTv2 & RT & 54.00 & 61.30 & 48.10 & 71.10 & 60.90 \\
UnisMOT~\cite{ref27} & PR & 2025 & CSPDarknet & RT & 54.20 & 59.50 & 49.70 & 65.70 & 60.30 \\
UnisMOT~\cite{ref27} & PR & 2025 & CSPDarknet & RNT & 56.60 & 62.20 & 51.70 & 69.30 & 62.90 \\
PFTrack~\cite{ref28} & PR & 2025 & DLA34 & RT & 53.94 & 55.25 & 53.22 & 66.64 & 65.55 \\
MCTrack~\cite{ref29} & TCSVT & 2026 & DLA34 & RT & 57.35 & 59.15 & 56.78 & 70.14 & 66.75 \\
FairMOT~\cite{ref1} & IJCV & 2021 & DLA34 & RT & 59.74 & 57.46 & \textbf{62.62} & 70.96 & \underline{74.75} \\
SATA~\cite{ref37} & AAAI & 2026 & Hiera-L & RT & 59.70 & \underline{63.60} & 53.70 & 75.10 & 65.40 \\

\rowcolor{lightblue}
Ours & -- & -- & DLAMM34 (ours) & RT & \underline{61.66} & 62.54 & 61.32 & \underline{75.84} & \underline{73.99} \\

\rowcolor{lightblue}
Ours & -- & -- & DLAMM34 (ours) & RNT & \textbf{63.31} & \textbf{64.93} & \underline{62.39} & \textbf{79.21} & \textbf{76.02} \\

\bottomrule
\end{tabular}
}
\end{table*}

As shown in Table~\ref{tab:sota_tracking}, under the RT modality, the proposed method achieves HOTA and MOTA scores of 61.66 and 75.84, respectively, outperforming the second-best method FairMOT by 1.92 and 4.88 percentage points. The IDF1 score reaches 73.99, indicating strong identity consistency while effectively reducing identity switches. It is worth noting that FairMOT achieves a higher association accuracy (AssA) of 62.62, which is 1.30 percentage points higher than our method. This is mainly because FairMOT uses appearance features during the data association stage, whereas our method employs a lightweight IoU-based tracker.

When NIR features are further introduced under the RNT setting, all performance metrics improve significantly. Specifically, HOTA, DetA, and MOTA increase to 63.31, 64.93, and 79.21, respectively. Compared with UnisMOT under the same RNT setting, the proposed method improves by 6.71, 2.73, and 10.69 percentage points in HOTA, DetA, and MOTA, respectively. This significant improvement fully demonstrates the effectiveness of the proposed architecture, and also verifies that multimodal tracking outperforms dual-modal tracking.

\subsubsection{Illumination Robustness Analysis}

To evaluate the robustness of the proposed method under complex lighting conditions, experiments are conducted on three subsets: low-illumination, middle-illumination, and high-illumination scenarios. Since PFTrack~\cite{ref28}, MCTrack~\cite{ref29}, and SATA~\cite{ref37} do not provide results under different illumination conditions in their original papers, they are excluded from this comparison. The results are shown in Table~\ref{tab:illumination}.

\begin{table*}[!t]
\centering
\caption{Tracking performance under low-, medium-, and high-illumination conditions}
\label{tab:illumination}
\renewcommand{\arraystretch}{1.15}

\resizebox{\textwidth}{!}{
\begin{tabular}{cccccccccccc}
\toprule
\multirow{2}{*}{Methods} & \multirow{2}{*}{Venue} & \multirow{2}{*}{Year} &
\multicolumn{3}{c}{Low-illumination} &
\multicolumn{3}{c}{Middle-illumination} &
\multicolumn{3}{c}{High-illumination} \\
\cmidrule(lr){4-6}
\cmidrule(lr){7-9}
\cmidrule(lr){10-12}
& & &
HOTA$\uparrow$ & DetA$\uparrow$ & AssA$\uparrow$ &
HOTA$\uparrow$ & DetA$\uparrow$ & AssA$\uparrow$ &
HOTA$\uparrow$ & DetA$\uparrow$ & AssA$\uparrow$ \\
\midrule

UniTrack~\cite{ref35} & NeurIPS & 2021 & 18.20 & 15.90 & 20.80 & 53.50 & 51.10 & 56.70 & 59.30 & 63.50 & 56.90 \\
Unicorn~\cite{ref36} & ECCV & 2022 & 21.70 & 19.20 & 24.80 & 56.30 & 54.30 & 58.90 & \underline{62.20} & \underline{65.70} & 59.60 \\
MOTRv2~\cite{ref3} & CVPR & 2023 & 24.50 & 22.10 & 27.50 & 57.80 & 58.70 & 64.20 & 55.63 & 49.60 & 62.40 \\
ByteTrack~\cite{ref2} & ECCV & 2022 & 33.50 & 36.90 & 30.70 & 58.70 & 60.90 & 56.90 & 60.70 & \textbf{65.80} & 56.40 \\
HGT-Track~\cite{ref26} & TMM & 2024 & 47.60 & 58.10 & 39.50 & 55.10 & 59.90 & 51.00 & 51.90 & 52.00 & 52.10 \\
UnisMOT~\cite{ref27} & PR & 2025 & 49.80 & \underline{58.70} & 42.50 & 54.40 & 55.20 & 53.80 & 56.60 & 56.20 & 57.30 \\
FairMOT~\cite{ref1} & IJCV & 2021 & \underline{53.24} & 58.58 & \textbf{48.85} & \underline{63.30} & \underline{61.72} & \underline{65.31} & 58.48 & 52.58 & \textbf{65.64} \\

\rowcolor{lightblue}
Ours & -- & -- & \textbf{55.32} & \textbf{64.81} & \underline{47.84} & \textbf{66.42} & \textbf{67.36} & \textbf{65.99} & \textbf{63.06} & 62.02 & \underline{64.83} \\

\bottomrule
\end{tabular}
}
\end{table*}

As shown in Table~\ref{tab:illumination}, under low-illumination conditions, the performance of traditional single-modal or weak fusion methods degrades significantly. In contrast, the proposed method achieves the best HOTA and DetA scores, improving by 2.08 and 6.23 percentage points over FairMOT, respectively. This demonstrates that multimodal fusion provides strong complementary information under poor visibility conditions.

Under middle-illumination conditions, the proposed method achieves the best performance in HOTA, DetA, and AssA, outperforming FairMOT by 3.12, 5.64, and 0.68 percentage points, respectively. FairMOT slightly outperforms in AssA in some cases due to its strong appearance feature modeling capability, but the proposed method still achieves the highest overall performance.

Under high-illumination conditions, ByteTrack achieves the highest DetA (65.80), and FairMOT achieves the highest AssA (65.64), both slightly outperforming the proposed method. This indicates that in well-lit conditions, redundant information from NIR/TIR may introduce slight interference to RGB features. Nevertheless, the proposed method still maintains competitive tracking performance in this scenario.

Overall, the proposed method demonstrates strong robustness in challenging low- and medium-illumination environments. Although performance slightly decreases under ideal lighting conditions due to multimodal redundancy, its overall performance across diverse conditions verifies its strong adaptability. This further demonstrates that the proposed amplitude–phase decoupled spatial–modal fusion mechanism effectively enhances cross-modal structural consistency, improving perception and association stability under weak illumination, occlusion, and modality degradation.

\subsubsection{Efficiency Analysis}

To evaluate inference efficiency, experiments are conducted under unified hardware and inference settings. Since some compared methods do not report GFLOPs or parameter counts, four representative methods with publicly available efficiency metrics are selected for comparison. The results are shown in Table~\ref{tab:efficiency}.

\begin{table*}[!t]
\centering
\caption{Comparison of inference efficiency for different tracking models}
\label{tab:efficiency}
\renewcommand{\arraystretch}{1.15}

\resizebox{\textwidth}{!}{
\begin{tabular}{cccccccccc}
\toprule
Methods & Venue & Year & Backbone & HOTA$\uparrow$ & DetA$\uparrow$ & AssA$\uparrow$ & Params$\downarrow$ & GFLOPs$\downarrow$ & FPS$\uparrow$ \\
\midrule

ByteTrack~\cite{ref2} & ECCV & 2022 & CSPDarknet & 53.40 & 58.20 & 49.60 & 41.30M & \textbf{115.82} & \textbf{34.10} \\
HGT-Track~\cite{ref26} & TMM & 2024 & PVTv2 & 54.00 & \underline{61.30} & 48.10 & 30.70M & 145.51 & 15.21 \\
UnisMOT~\cite{ref27} & PR & 2025 & CSPDarknet & 54.20 & 59.50 & 49.70 & 99.15M & 217.46 & 24.80 \\
FairMOT~\cite{ref1} & IJCV & 2021 & DLA34 & \underline{59.74} & 57.46 & \textbf{62.62} & \underline{19.66M} & 167.96 & 29.74 \\

\rowcolor{lightblue}
Ours & -- & -- & DLAMM34 (ours) & \textbf{63.31} & \textbf{64.93} & \underline{62.39} & \textbf{14.59M} & \underline{139.69} & \underline{31.55} \\

\bottomrule
\end{tabular}
}
\end{table*}

ByteTrack achieves 34.10 FPS with a lightweight detector, but its HOTA is only 53.40 and AssA is below 50. In contrast, the proposed method introduces lightweight Basic/Mixed convolution modules in the backbone, reducing parameters to 14.59M and computational cost to 139.69 GFLOPs, while still achieving the best HOTA and DetA. Meanwhile, it maintains a real-time speed of 31.55 FPS, demonstrating a strong balance between accuracy and efficiency.

\subsection{Visualization Analysis}

To intuitively validate the effectiveness of the proposed method under multimodal fusion and challenging illumination conditions, this section presents a comprehensive visualization analysis from three perspectives: feature representations, modality weights, and final tracking results.

\subsubsection{Amplitude–Phase Feature Visualization}

To further evaluate the cross-modal modeling capability of the MS-Mixed convolution, low-illumination scenes are selected to visualize the outputs of different Tree branches at Level 4 and Level 5 of the network. Specifically, Tree 1-1 at Level 4 denotes the first sub-branch of the first branch in the fourth layer, while Tree 1 at Level 5 denotes the first branch of the fifth layer. The results are shown in Fig.~12. Fig.~12(a) illustrates the amplitude component after convolution and channel-wise averaging, where red indicates high response values. Fig.~12(b) shows the phase component after convolution and ReLU activation, where red indicates high activation values.

The input contains three modalities, i.e., RGB, NIR, and TIR. Under low illumination conditions, the RGB channel is nearly black due to severe degradation, while the NIR modality preserves clear pedestrian contours, and the TIR modality exhibits strong thermal radiation responses but lacks fine texture information.

\begin{figure*}[!t]
\centering
\includegraphics[width=\textwidth]{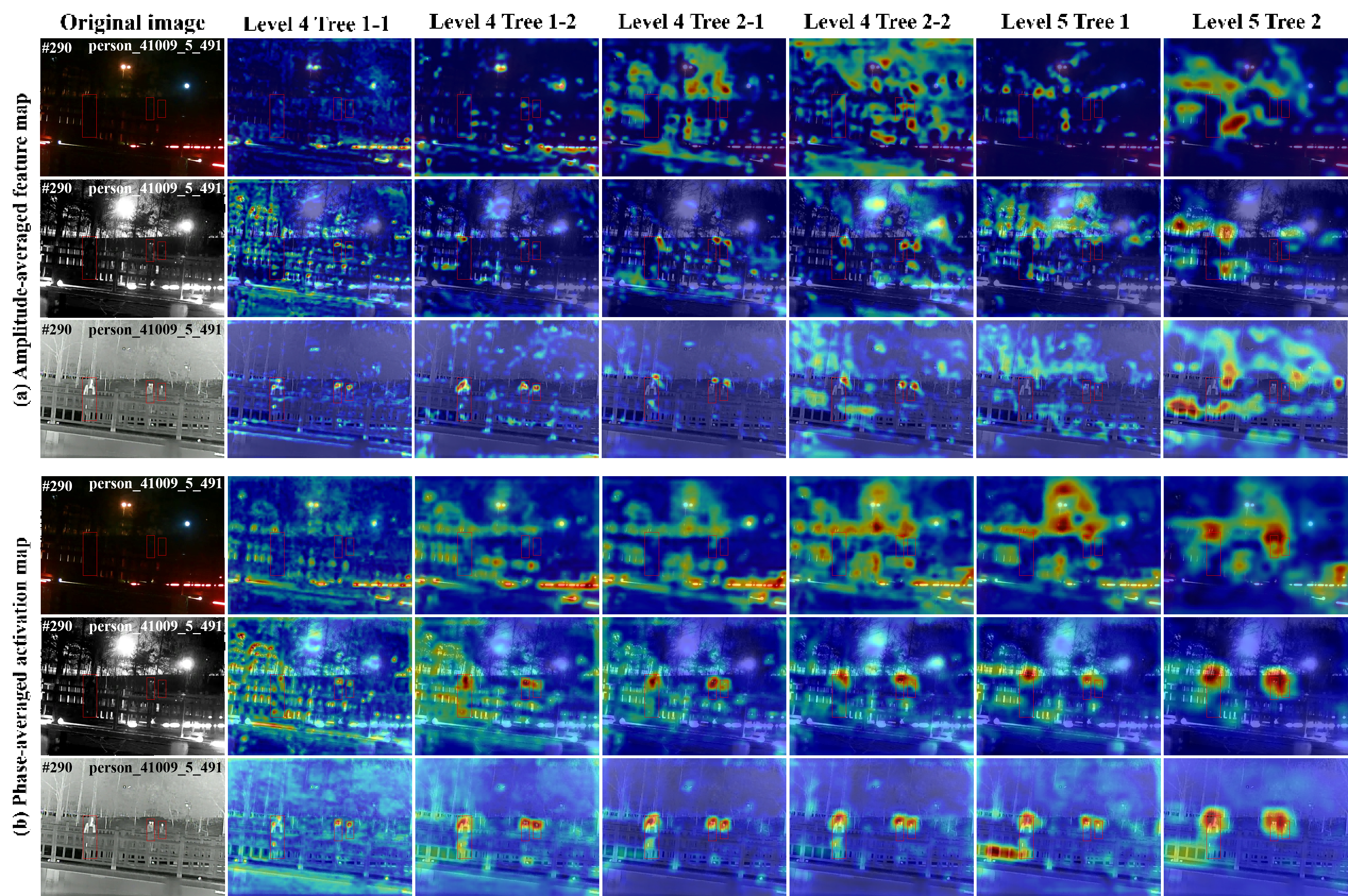}
\caption{Visualization of amplitude and phase in MS-Mixed convolution.}
\label{fig12}
\end{figure*}

The input contains RGB, NIR, and TIR modalities. Under low illumination, RGB responses are severely degraded, whereas NIR preserves clearer structural contours and TIR provides strong thermal saliency.

As shown in Fig.~\ref{fig12}, the amplitude branch mainly enhances target-related semantic responses, where NIR and TIR exhibit significantly stronger activations than RGB. In contrast, the phase branch maintains stable structural responses across modalities while suppressing background interference, demonstrating its capability in modeling cross-modal structural consistency.

Furthermore, regions with similar structural patterns across different modalities exhibit increasingly consistent activations in the phase branch, whereas modality-specific response discrepancies are gradually weakened. This behavior suggests that the proposed complex-valued amplitude--phase modeling implicitly promotes modality-consistent structural information while suppressing inconsistent cross-modal responses. As a result, the network can better preserve shared target topology and alleviate modality interference during deep semantic fusion.

These results indicate that the proposed Mixed module effectively achieves joint modeling of semantic saliency and structural relationships through amplitude--phase decomposition.

\subsubsection{Modality Weight Analysis in DPG}

To evaluate the illumination-adaptive capability of the proposed Distillation Prompt-Guided (DPG) network, modality weights under high-, medium-, and low-illumination conditions are visualized in Fig.~13--Fig.~15. The outputs are normalized by Softmax and illustrated using radar plots. Since Level 4 exhibits nearly uniform modality responses, only representative results from Levels 1, 2, 3, and 5 are presented.

\begin{figure}[!t]
\centering
\includegraphics[width=\columnwidth]{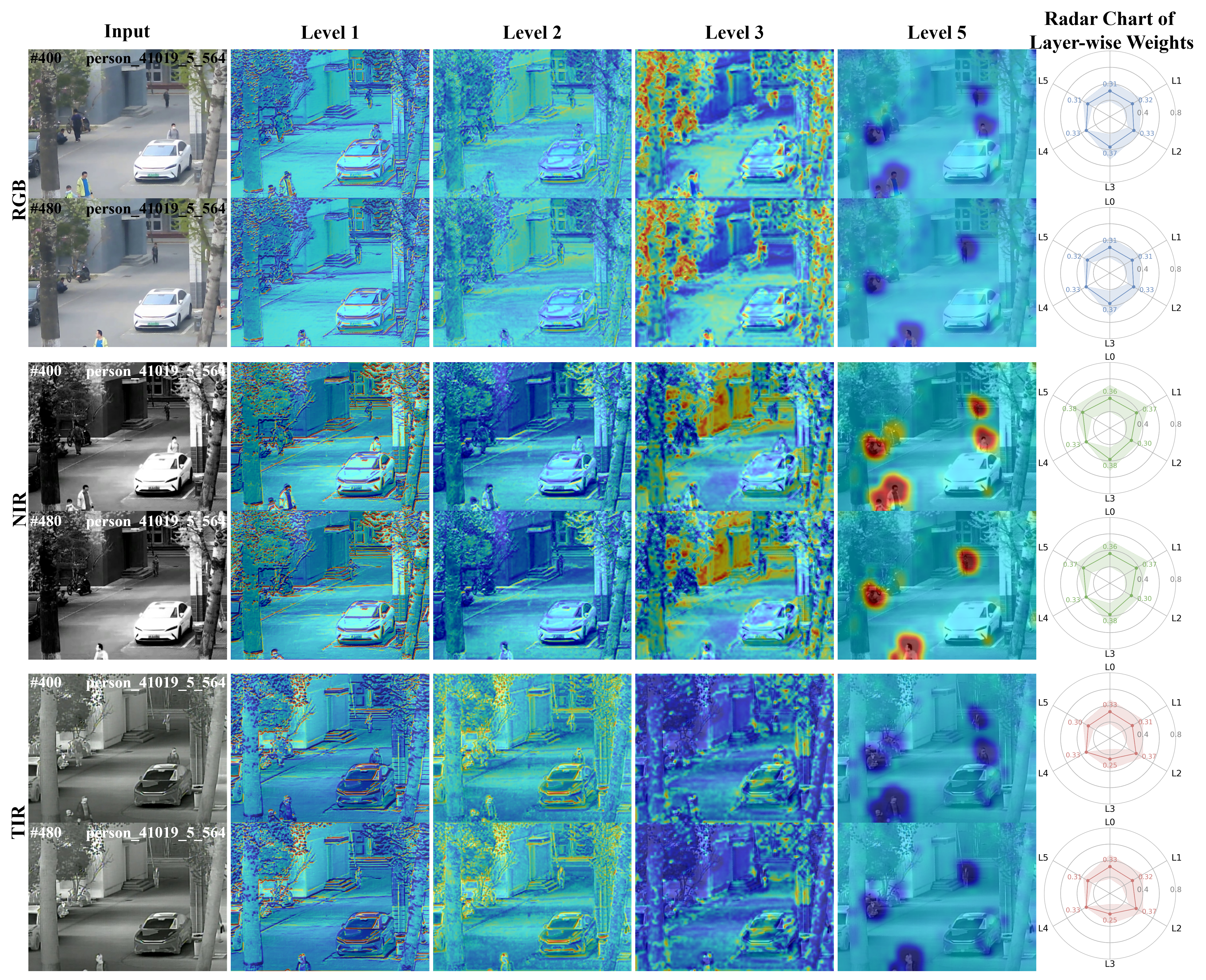}
\caption{Modality weight visualization under high-illumination conditions.}
\label{fig13}
\end{figure}

\begin{figure}[!t]
\centering
\includegraphics[width=\columnwidth]{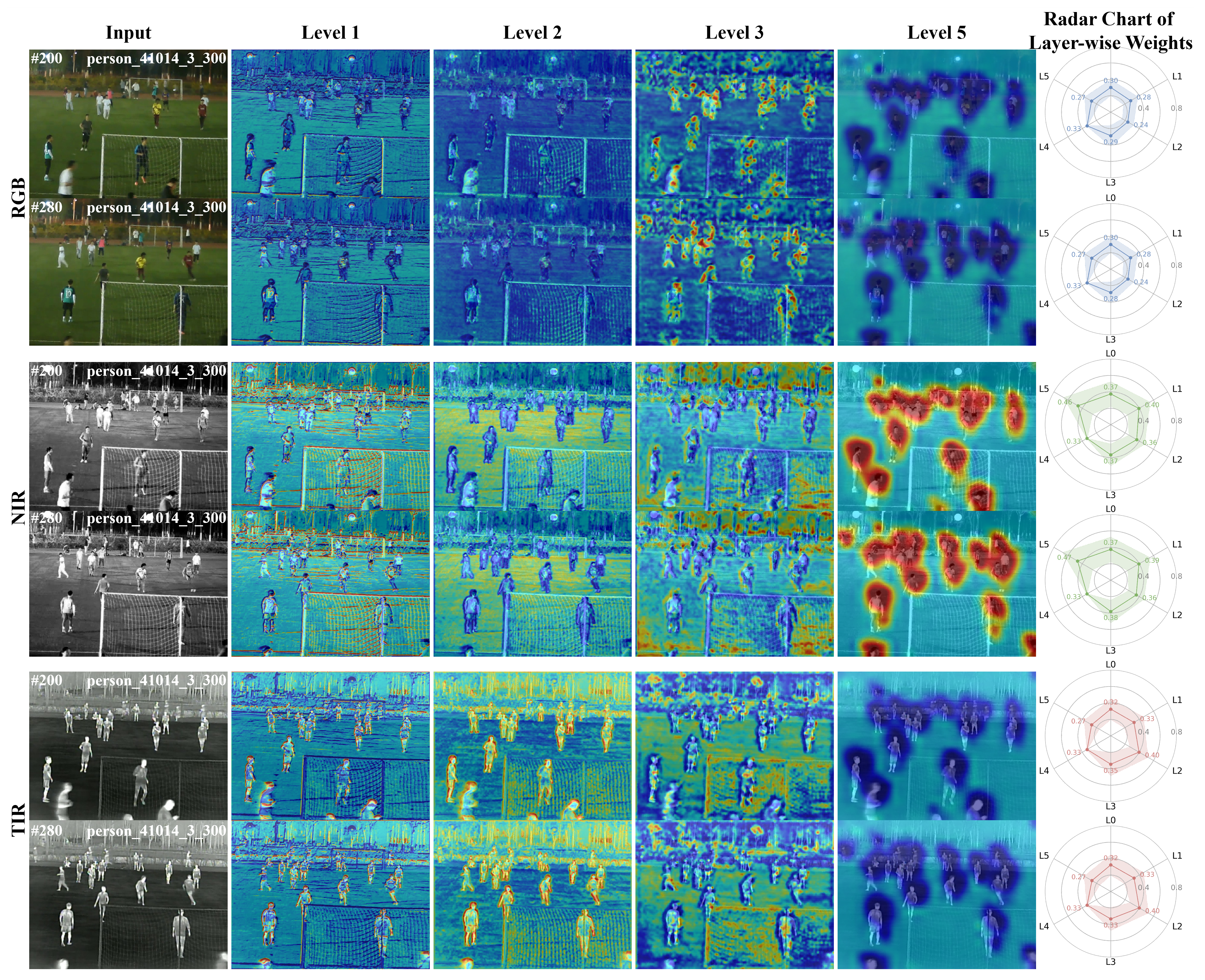}
\caption{Modality weight visualization under medium-illumination conditions.}
\label{fig14}
\end{figure}

\begin{figure}[!t]
\centering
\includegraphics[width=\columnwidth]{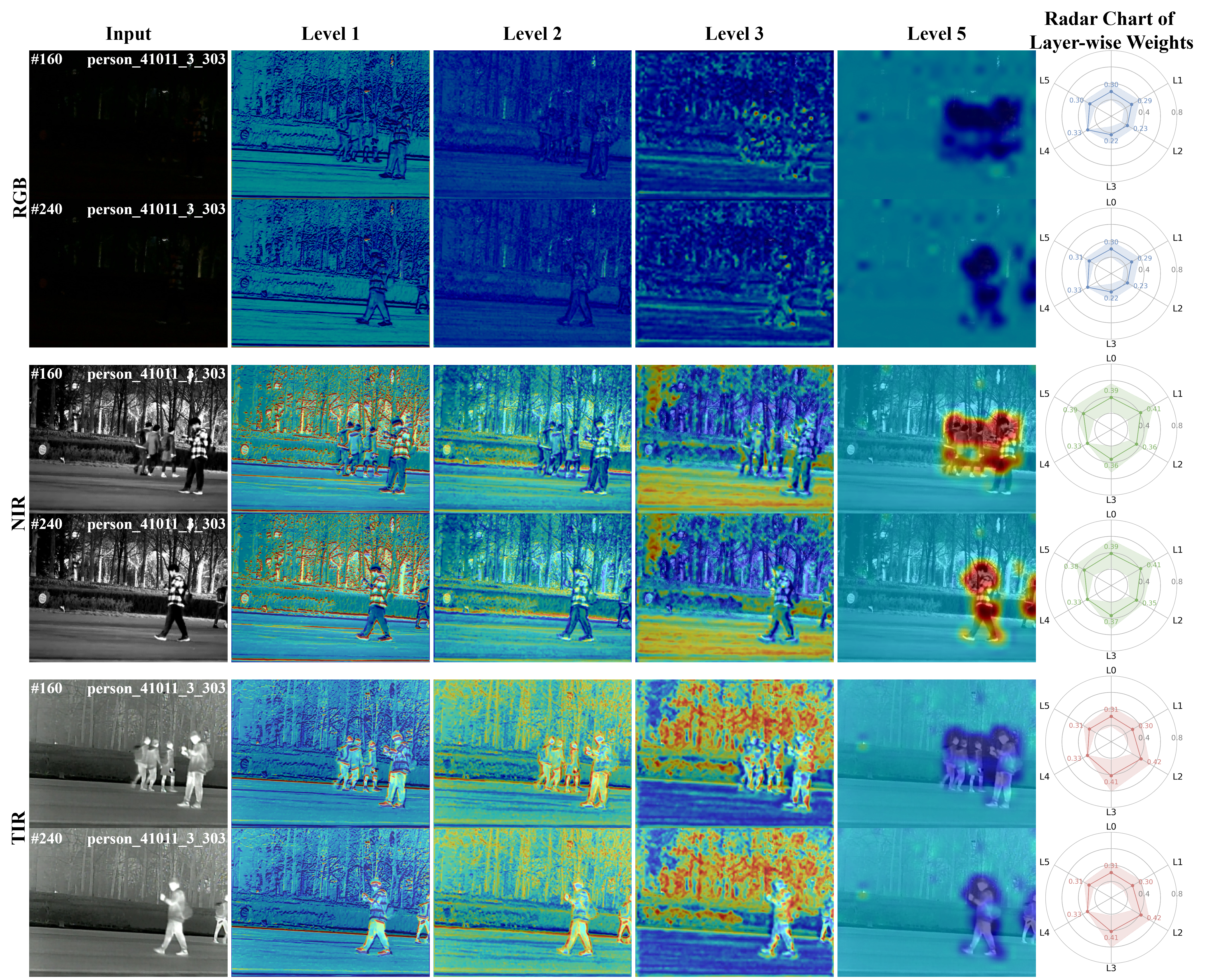}
\caption{Modality weight visualization under low-illumination conditions.}
\label{fig15}
\end{figure}

Under high-illumination conditions, RGB and NIR dominate the fusion process, while TIR only contributes in several intermediate layers. The modality distribution remains relatively balanced, indicating that visible-spectrum information is sufficiently preserved under strong illumination.

Under medium illumination, the contribution of RGB decreases noticeably, whereas NIR and TIR responses increase progressively, demonstrating that the network gradually shifts toward infrared modalities as illumination degrades.

Under low-illumination conditions, RGB responses become severely suppressed, and the fusion process mainly relies on NIR and TIR modalities, which provide more stable structural and thermal information in dark environments.

Overall, the proposed DPG network dynamically reallocates modality importance across different illumination conditions, enabling more robust and adaptive multimodal fusion.

\subsubsection{Qualitative Comparison with State-of-the-Art Methods}

To further evaluate the robustness of the proposed method, qualitative comparisons with FairMOT, MCTrack, PFTrack, and UnisMOT are conducted under crowded, low-illumination, fast-motion, and overexposed scenarios, as shown in Fig.~16. Here, M and S denote missed detections and identity switches, respectively.

\begin{figure*}[!t]
\centering
\includegraphics[width=\textwidth]{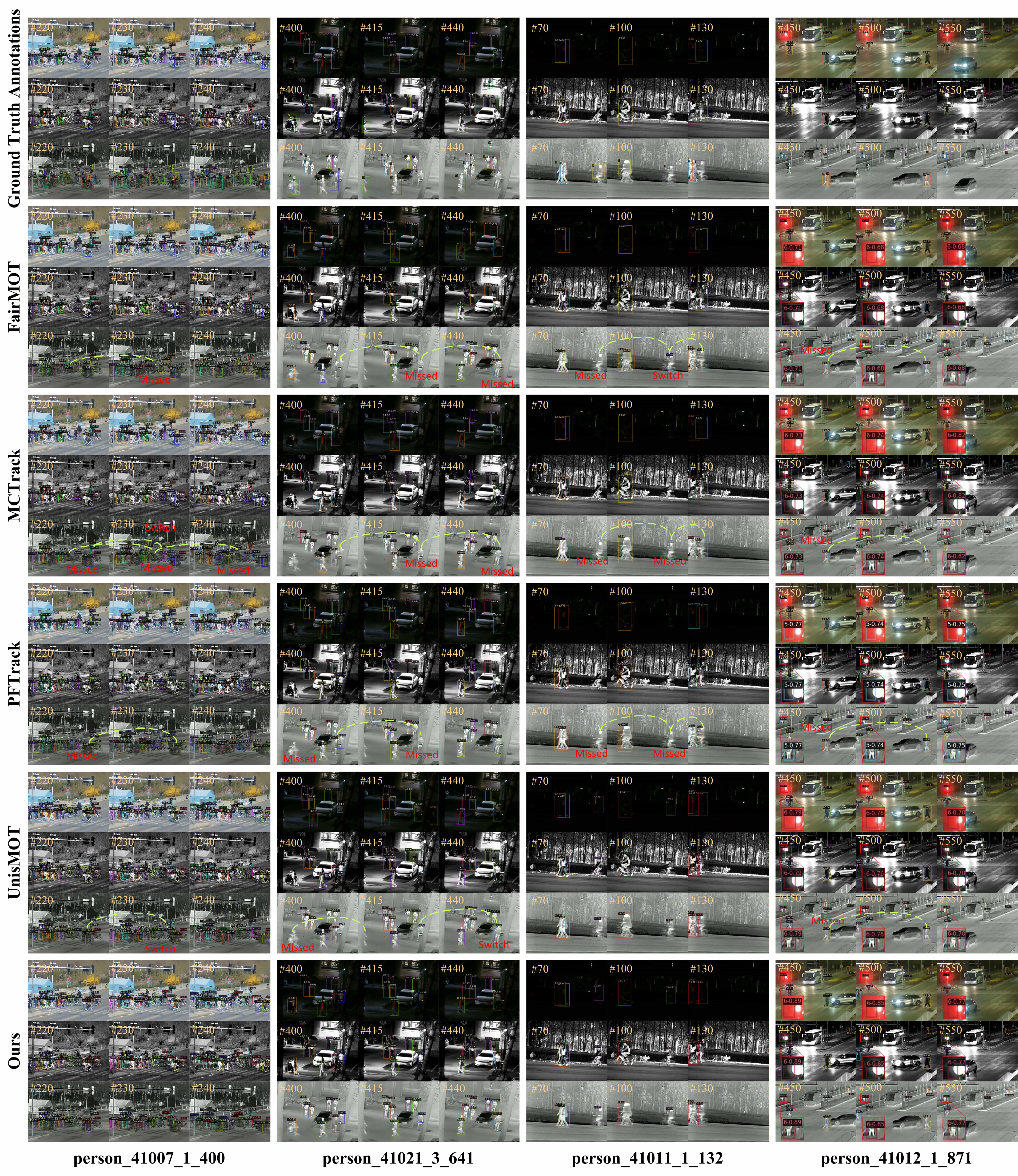}
\caption{Visualization comparison of different methods under complex scenarios.}
\label{fig16}
\end{figure*}

In crowded scenes (first column), most competing methods suffer from missed detections or identity switches due to heavy occlusion and dense interactions. In contrast, the proposed method maintains stable identity association, benefiting from the enhanced spatial-modal representation capability of the Basic and Mixed modules.

In low-illumination scenes (second column), conventional methods exhibit unstable tracking due to severe RGB degradation. By adaptively increasing the contribution of NIR and TIR modalities, the proposed method maintains more reliable target localization and identity consistency under poor lighting conditions.

In fast-motion scenarios (third column), motion blur and short-term occlusion lead to trajectory fragmentation in competing methods. In contrast, the proposed method preserves more stable trajectories by jointly modeling response saliency and cross-modal structural consistency.

In overexposed scenes (fourth column), severe RGB saturation causes detection failures in most competing methods, whereas the proposed method still maintains stable target localization and trajectory association by relying on complementary infrared information.

Overall, the qualitative results demonstrate that the proposed method achieves more robust detection and identity preservation under challenging conditions, validating the effectiveness of the proposed multimodal fusion and adaptive representation mechanisms.

\section{Conclusion and Future Work}

This paper addresses two key challenges in multimodal multi-object tracking: the insufficient joint modeling of spatial and modality information, and the failure of fixed fusion strategies under drastic illumination changes.

To address these issues, a spatial-modal convolutional fusion backbone is first proposed. A shallow Basic module is designed to decouple spatial feature extraction and modality interaction via a stepwise 3D convolution scheme, avoiding early entanglement of fine-grained features. A deep Mixed module is further introduced, which employs an amplitude--phase decomposition mechanism to model complex nonlinear inter-modal relationships, enabling unified optimization across spatial and modality dimensions.

Secondly, a representation collapse network based on distillation prompts and global aggregation is constructed. A DPG network is designed to guide a student network using a teacher network, enabling adaptive generation of modality fusion weights according to illumination conditions. In addition, a GMDA module is proposed to jointly model low-level modality commonality and high-level cross-modal differences, enabling high-quality representation collapse from 3D multimodal features into compact 2D embeddings.

Extensive ablation studies validate the effectiveness of each component. Comparative experiments demonstrate that the proposed method achieves the best performance across all tracking metrics while maintaining a favorable balance between accuracy and efficiency. Amplitude--phase visualization confirms the effective modulation of target responses and cross-modal discrepancies. Modality weight visualization verifies the effectiveness of adaptive reweighting under varying illumination conditions. Qualitative results further demonstrate that the proposed method maintains robust detection and identity consistency under highly dynamic scenarios, including illumination variation, occlusion, and motion blur, demonstrating strong applicability for complex multi-object tracking tasks.

\bibliographystyle{IEEEtran}
\bibliography{references}

\end{document}